\documentclass[aps,pre,showpacs,reprint,onecolumn,groupedaddress]{revtex4-1}
\usepackage[utf8]{inputenc}
\usepackage{amsmath}
\usepackage{graphicx}
\usepackage{bbm}
\usepackage{amsfonts}
\usepackage{array}
\usepackage{algorithm}
\usepackage{algorithmic}
\usepackage[caption=false]{subfig}
\usepackage{accents}
\newlength{\dhatheight}

\newcommand{\iid}{i.i.d. }
\newcommand{\fh}{\hat{f}}
\newcommand{\fb}{\bar{f}}
\newcommand{\xh}{\hat{x}}
\newcommand{\xb}{\bar{x}}

\newcommand{\NN}{\mathcal{N}}

\newcommand{\RR}{\mathbb{R}}
\newcommand{\ZZ}{\mathcal{Z}}
\newcommand{\p}{p}

\newcommand{\dd}{{\rm d}}
\newcommand{\DD}[1]{{\rm D} #1 \,}

\newcommand{\Q}{\mathbf{Q}}
\newcommand{\Qh}{\hat{\mathbf{Q}}}
\newcommand{\Qz}{\mathbf{Q}_z}
\newcommand{\Qu}{\mathbf{Q}_u}
\newcommand{\Qv}{\mathbf{Q}_v}

\newcommand{\Qhu}{\hat{\mathbf{Q}}_u}
\newcommand{\Qhv}{\hat{\mathbf{Q}}_v}
\newcommand{\ua}{\mathbf{u}}
\newcommand{\va}{\mathbf{v}}
\newcommand{\z}{\mathbf{z}}
\newcommand{\Zh}{\hat{Z}} \newcommand{\Zb}{\bar{Z}}  	\newcommand{\Zv}{\mathbf{Z}}	\newcommand{\zv}{\mathbf{z}}
\newcommand{\Zhv}{\hat{\mathbf{Z}}} \newcommand{\Zbv}{\bar{\mathbf{Z}}} 	 
\newcommand{\Tr}{{\rm Tr}}
\newcommand{\ac}{S}
\newcommand{\f}{f}
\newcommand{\lr}{\mathbf{X}}  \newcommand{\X}{\mathbf{X}}
\newcommand{\U}{U} \newcommand{\Uv}{\mathbf{U}}  \newcommand{\us}{\mathbf{u}}  
\newcommand{\Uh}{\hat{U}}	\newcommand{\Uhv}{\hat{\mathbf{U}}}
\newcommand{\Ub}{\bar{U}}	\newcommand{\Ubv}{\bar{\mathbf{U}}}
\newcommand{\uh}{\hat{u}}	\newcommand{\uhv}{\hat{\mathbf{u}}}
\newcommand{\ub}{\bar{u}}	\newcommand{\ubv}{\bar{\mathbf{u}}}
  	\newcommand{\Vv}{\mathbf{V}}	\newcommand{\vs}{\mathbf{v}}
\newcommand{\Vh}{\hat{V}}	\newcommand{\Vhv}{\hat{\mathbf{V}}}
\newcommand{\Vb}{\bar{V}}	\newcommand{\Vbv}{\bar{\mathbf{V}}}
\newcommand{\vh}{\hat{v}}	\newcommand{\vhv}{\hat{\mathbf{v}}}
\newcommand{\vb}{\bar{v}}	\newcommand{\vbv}{\bar{\mathbf{v}}}
\newcommand{\A}{\mathcal{A}}  \newcommand{\Au}{\mathcal{A}_U}  \newcommand{\Av}{\mathcal{A}_V}
\newcommand{\AM}{\mathbf{A}}	\newcommand{\as}{A}
\newcommand{\Z}{\mathbf{Z}}
\newcommand{\y}{\mathbf{Y}} \newcommand{\yv}{\mathbf{Y}}
\newcommand{\rank}{R} \newcommand{\ri}{s}  \newcommand{\ribis}{s'}
\newcommand{\US}{M}  \newcommand{\ui}{\mu} \newcommand{\uibis}{\mu'}
\newcommand{\VS}{P}  \newcommand{\vi}{p}   \newcommand{\vibis}{p'}
\newcommand{\YS}{L}  \newcommand{\yi}{l}	\newcommand{\yibis}{l'}
\newcommand{\gh}{\hat{g}} 	\newcommand{\ghv}{\mathbf{\hat{g}}}
\newcommand{\gb}{\bar{g}}	\newcommand{\gbv}{\mathbf{\bar{g}}}
\newcommand{\g}{\mathbf{g}}
\newcommand{\iu}{\mathcal{I}_U}
\newcommand{\iv}{\mathcal{I}_V}

\newcommand{\iz}{\mathcal{I}_Z}
\newcommand{\au}{\alpha_U} \newcommand{\av}{\alpha_V} 
\newcommand{\iter}{i}
\newcommand{\smu}{Q_u^0} \newcommand{\smv}{Q_v^0}
\newcommand{\msex}{{\rm MSE}_x} \newcommand{\mseu}{{\rm MSE}_u} \newcommand{\msev}{{\rm MSE}_v}
\newcommand{\messu}{m} \newcommand{\messuh}{\tilde{m}} \newcommand{\messv}{n} \newcommand{\messvh}{\tilde{n}}
\newcommand{\pbig}{P-BiG-AMP}

\DeclareMathOperator{\extr}{{\rm extr}}

\begin{document}
\title{Phase diagram of matrix compressed sensing}

\author{Christophe Sch\"ulke}
\email{christophe.schulke@espci.fr}
 \affiliation{Laboratoire de Physique Statistique, CNRS, PSL Universit\'es et Ecole Normale Sup\'erieure, 75005, Paris, France}
\affiliation{Institut de Physique Théorique, CNRS, CEA, Université Paris-Saclay, 91191, Gif-sur-Yvette, France}

\author{Philip Schniter}
\email{schniter@ece.osu.edu}
\affiliation{Department of Electrical and Computer Engineering, The Ohio State University, Columbus, OH 43210 ,USA }

\author{Lenka Zdeborov\'a}
\email{lenka.zdeborova@cea.fr}
\affiliation{Institut de Physique Théorique, CNRS, CEA, Université Paris-Saclay, 91191, Gif-sur-Yvette, France}

\begin{abstract}
 In the problem of matrix compressed sensing we aim to recover a
 low-rank matrix from few of its element-wise linear projections. In
 this contribution we analyze the asymptotic performance of a
 Bayes-optimal inference procedure for a model where the matrix to be
 recovered is a product of random matrices. The results that we
 obtain using the replica method describe the state evolution of
 the recently introduced P-BiG-AMP \cite{parker2015parametric} algorithm. 
 We show the existence of different types of phase transitions, their implications for the solvability of the problem, 
 and we compare the results of the theoretical analysis to the
 performance reached by P-BiG-AMP. Remarkably the asymptotic replica
 equations for matrix compressed sensing are the same as those for a
 related but
 formally different problem of matrix factorization \cite{kabashima2014phase}.
\end{abstract}

\maketitle

\section{Introduction}

Recovering a sparse or a low-rank signal from as few observations as
possible is a class of problems that attracted considerable attention in
statistics and signal processing.  Very popular examples of problems belonging to
this class are compressed sensing \cite{donoho2006compressed},
or matrix completion \cite{candes2009exact}.  Another interesting
member of this class is the problem matrix compressed sensing, in which
one aims to recover a low-rank matrix 
from a few of its random component-wise linear  projections. We give a
formal definition of the problem in Sec.~\ref{definition}.  This
problem has a range of interesting applications, see
\cite{parker2015parametric} and references therein. 

The main line
of theoretical work related to matrix compressed sensing minimizes the nuclear norm
of the matrix (i.e. the sum of its singular values) subject
to the constraint that its linear projections agree with the measured
values \cite{nuclearNormMin,Donoho21052013}. Nuclear norm minimization
is algorithmically tractable and when analyzed it provably recovers
the unknown matrix for an interesting range of parameters. 
The nuclear norm is a common type of regularization that enhances
low-rank solutions. A rank $\rank$ matrix $\lr$ of dimension $\US \times \VS$  can
be written as a product of two matrices $\lr = \Uv \Vv^{\top}$ of sizes $\US \times \rank$ and
$\rank \times \VS$. However, the nuclear norm minimization approach
does not handle straightforwardly cases when there are some
requirements (such as sparsity)  on the factors
$\Uv$ and $\Vv$.


In the present paper we study the generalized matrix compressed sensing
problem, where arbitrary component-wise constraints are put on the
factors $\Uv$ and $\Vv$ and  each of the linear projections is observed trough some non-linear
output channel. We are able to do that if we restrict to a probabilistic model where the
components of the ground-truth factors $\Uv$ and $\Vv$ are \iid random
variables of known probability distribution, and where the
probabilistic nature of the output channel is known. Under such
assumptions the model is amenable to exact analysis via the replica
method developed in statistical physics \cite{mezard1990spin,nishimoriBook}. The results stemming from the
replica method are in general in one-to-one correspondence with the analysis of
message passing algorithms designed to solve the problem in an optimal
way, as illustrated for the compressed sensing problem in \cite{krzakaCS} or
for matrix factorization in \cite{kabashima2014phase}. For the matrix
compressed sensing problem this algorithm, called \pbig, was
derived and tested recently in \cite{parker2015parametric}. Our contribution can hence
also be viewed as an asymptotic analysis of the performance of this
algorithm for the assumed model. We
compare the analysis to the performance of \pbig\, and indeed observe excellent
agreement.

Our analysis reveals a striking connection between the matrix
compressed sensing problem and the problem of matrix factorization as
studied in \cite{krzakala2013phase,kabaSample,kabashima2014phase}. These are two different
inference problems. In matrix compressed sensing we observe a set of
element-wise linear projections of the matrix, whereas in matrix
factorization we observe the elements of the matrix directly. Yet the
replica analysis of the two problems yields equivalent equations and
hence the asymptotic behavior of the two problems, including the
phase transition, is closely linked. This analogy was already
remarked for the nuclear norm minimization for matrix compressed
sensing and matrix denoising in \cite{Donoho21052013}, or for matrix
compressed sensing and matrix completion in~\cite{Bolcskei}. 

\subsection{Definition of the problem}
\label{definition}
\begin{figure}[h]
 \centering
 \includegraphics[width=0.8\columnwidth]{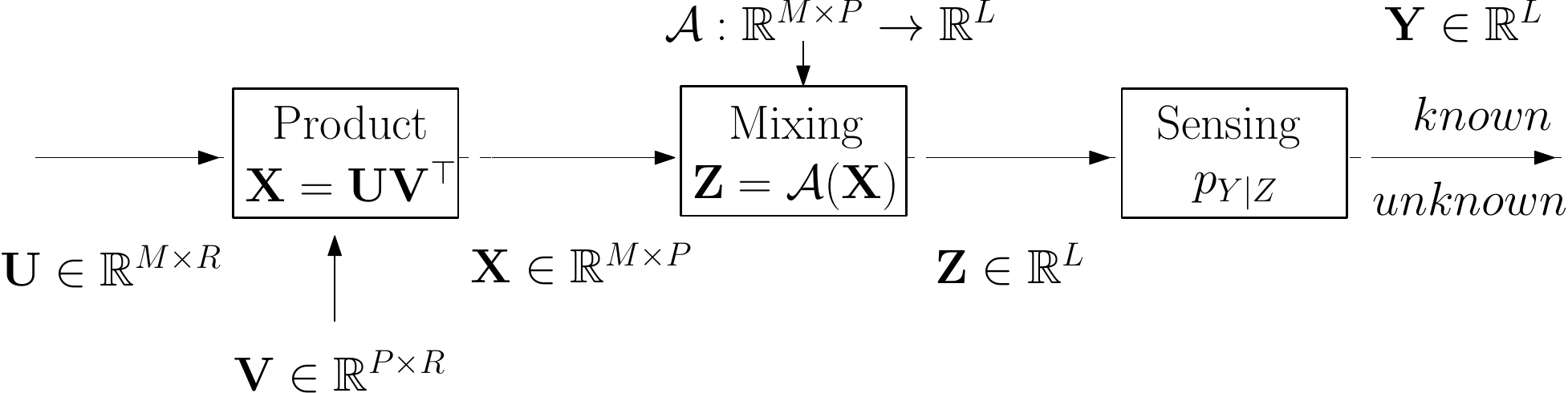}
 \caption{The setting of generalized matrix compressed sensing. 
 A low-rank matrix $\lr$ can be decomposed into a product of two smaller matrices $\Uv$ and $\Vv$.
 A linear operator $\A$ is applied to $\lr$, producing an intermediary variable $\Z$. 
 A measurement $\y$ of $\Z$ is obtained through a noisy channel. 
 The problem is closely linked to other inference problems: dropping the ``mixing'' block, one recovers 
 a generalized matrix factorization problem. Dropping the ``product''
 block, one recovers a generalized linear model.
}
 \label{fig:setting}
\end{figure}

Let $\lr \in \RR^{\US \times \VS}$ be a matrix of low rank $\rank< \min(\US, \VS)$. 
It can thus be written as a product of two smaller matrices: $\Uv \in \RR^{\US \times \rank}$ and $\Vv \in \RR^{\VS \times \rank}$,
\begin{equation}
 \lr = \Uv \Vv^{\top}.
\end{equation}
The low-rank matrix compressed sensing problem consists in recovering $\lr$ from a set of linear combinations of its entries.
We call $\A: \RR^{\US \times \VS} \to \RR^{\YS}$ the linear operator associated to the matrix $\AM$, we note
\begin{align}
 \Z &= \A(\lr) \in \RR^{\YS}		\label{eq:mixing}
\end{align}
and $\y$ the measured version of $\Z$ after passing through an
element-wise measurement channel:
\begin{align}
\y &\sim \p^0_{Y|Z}(\y|\Z). 		\label{eq:outputChannel}
\end{align}
This setting is shown in Fig.~\ref{fig:setting}, and the goal is to reconstruct $\Uv$ and $\Vv$ (but sometimes only $\lr$) from the knowledge of $\y$.

We can rewrite~(\ref{eq:mixing}) in the component-wise manner
\begin{align}
 \forall \yi \in [1,\YS], \quad z_{\yi} &= \sum_{\ui=1}^{\US} \sum_{\vi=1}^{\VS} \as_{\yi}^{\ui \vi} x_{\ui \vi}  .  \label{eq:mixingComponents}
\end{align}

\subsubsection{The probabilistic model and assumptions of our analysis.}
In order to enable the asymptotic analysis (i.e. when $\US,\VS, \YS \to \infty$) via the replica method we
introduce the following probabilistic model for matrix compressed
sensing.

\begin{itemize}
 \item We assume that elements of $\Uv$ and $\Vv$ are sampled
   independently at random such that 
 \begin{align}
\Uv &\sim \prod_{\mu s} \p^0_U(u_{\mu s}), & \Vv &\sim \prod_{ps} \p^0_V(v_{ps} ).    	\label{eq:priors0}
\end{align}
We assume the distributions $\p^0_U$ and $\p^0_V$ to have zero mean and respective variances $\smu$ and $\smv$ of order one.
 These distributions might not be known exactly: instead, we use zero-mean priors $\p_U$ and $\p_V$ believed to be close to $\p^0_U$ and $\p^0_V$ (in terms of Kullback-Leibler divergence).
 \item We assume the output distribution $\p^0_{Y|Z}$ to be separable  
 \begin{align}
     \p^0_{Y|Z} = \prod_{l} p^0_{Y|Z}(y_l|z_l)\, .
\end{align}
In the inference we use a distribution $\p_{Y|Z}$ we believe to be close to $\p^0_{Y|Z}$ (in terms of Kullback-Leibler divergence).
 \item We assume the matrix $\AM$ of the linear operator $\A$ to have normally distributed \iid elements 
 with zero mean and variance $1/(\rank \US \VS)$, such that the elements of $\Z$ have zero mean and variance $\smu \smv$. 
 This is the same assumption as is often made in compressed sensing, and differentiates the problem from matrix factorization, 
 in which $\A$ is the identity.
 \item We assume the dimensions $\US$, $\VS$ and $\YS$ to be large,
   but their following ratios to be of order one 
 \begin{align}
  \au &= \frac{\YS}{\rank \US}, & \av &= \frac{\YS}{\rank \VS}. \label{eq:alphas} 
 \end{align}
On the other hand, $\rank$ can be small. 

\end{itemize}

\subsubsection{Measures of recovery}
Given the estimates $(\Uhv,\Vhv,\hat{\lr})$ that an algorithm returns for $(\Uv,\Vv,\lr)$, 
the following mean squared errors quantify how close the estimates are from the real values:
\begin{align}
\mseu&=  \frac{\| \Uv - \Uhv \|_F^2}{ \US \rank}, & \msev&= \frac{ \| \Vv - \Vhv \|_F^2}{ \VS \rank}, & \msex&= \frac{ \| \lr - \hat{\lr} \|_F^2}{ \YS \rank}, \label{eq:mses}
\end{align}
where $|| \cdot ||_F$ is the Frobenius norm of a matrix.
Note that as in matrix factorization, there is an inherent ill-posedness when it comes to recovering the couple $(\Uv,\Vv)$. 
As a matter of fact, for any $\rank \times \rank$ invertible matrix $\mathbf{C}$, the couple $(\Uv \mathbf{C} , \Vv \left(\mathbf{C}^{-1} \right)^{\top})$ generates the same $\lr$ as $(\Uv,\Vv)$.
In some case, this ill-posedness can be lifted thanks to the distributions $p^0_U$ and $p^0_V$, but this is not always the case and might nevertheless be cause of trouble.
In that case, it is possible to have a very low $\msex$ but high $\mseu$ and $\msev$. 


In the setting where $\rank=1$, $\Uv$ and $\Uhv$ are vectors and we can consider the following definitions of normalized mean squared errors
\begin{align}
 {\rm nMSE}_u &= 1 - \frac{\left| \Uv^{\top} \Uhv \right| }{||\Uv||_2 ||\Uhv||_2} ,  & {\rm nMSE}_v &=  1 - \frac{\left| \Vv^{\top} \Vhv \right| }{||\Vv||_2 ||\Vhv||_2} \label{eq:nmses}
\end{align}
that take values between 0 and 1 and take into account all invariances of the problem: an nMSE of 0 indicates perfect reconstruction up to the scaling invariance.

\subsection{Notations}
We use bold letters for vectors and matrices and non-bold letters for scalars.
The elements of a vector $\mathbf{x}$ are noted $[\mathbf{x}]_i$ or $x_i$.
The operator $\odot$ is used for element-wise multiplication of vectors or matrices. 
$\mathbf{x}^{-1}$,  $\mathbf{x}^2$ and $\mathbf{x}^{\top}$ refer respectively 
to the component-wise inverse, the component-wise square and the transpose of the vector (or matrix) $\mathbf{x}$. 
If $\A$ is a linear operator and $\AM$ its matrix, we write $\A^2$ for the linear operator associated to $\AM^2$.
We use the notation $\imath \equiv \sqrt{-1}$.
Estimators $\hat{X}$ and $\hat{x}$ of a variable $X$ are the minimal mean squared error (MMSE) estimators 
of estimated probability distribution functions $\hat{P}(x)$ and $\hat{p}(x)$.
We note $\bar{X}$ and $\bar{x}$ the variances of these distributions and refer to them 
as \textit{uncertainties}, as they are a measure of the 
uncertainty of the estimators $\hat{X}$ and $\hat{x}$.

Using the matrix $\AM$, we can define two auxiliary linear operators $\Au: \RR^{\VS} \to \RR^{\YS \times \US}$ and 
$\Av: \RR^{\US} \to \RR^{\YS \times \VS}$ such that
\begin{align}
 [\Au(\vs)]_{\yi \ui} &\equiv \sum_{\vi} \as_{\yi}^{\ui \vi} v_{\vi},   \label{eq:Au} \\
 [\Av(\us)]_{\yi \vi} &\equiv \sum_{\ui} \as_{\yi}^{\ui \vi} u_{\ui}.	\label{eq:Av}
\end{align}

We note $x \sim \p_X(x)$ a random variable $x$ following the probability distribution $\p_X$. 
This holds also for vectors and matrices: $\mathbf{x} \sim \p_X(\mathbf{x})$. 
In that case, we say that $\p_X(\mathbf{x})$ is separable if each component $x_i$ of $\mathbf{x}$ 
is sampled independently from the others: $\forall i, \, x_i \sim \p_{X_i}(x_i)$, 
which we will note $\p_X$ as well if the components are identically distributed.

We write $f(x) \propto g(x)$ when the functions $f$ and $g$ are equal up to a multiplying constant that does not depend on $x$.
We write $K = O(1)$ (respectively $K= O(M)$) in order to signify that $K$ is of order $1$ (respectively $M$). 

Let us introduce some useful functions that will be used throughout the paper.
We note $\NN(x;\xh,\xb)$ the normalized Gaussian with mean $\xh$ and variance $\xb$:
\begin{equation}
 \NN(x;\xh,\xb) = \frac{1}{\sqrt{2 \pi \xb}} e^{-\frac{(x-\xh)^2}{2 \xb}}.	\label{eq:gaussian}
\end{equation}
In integrals, we note $\DD t$ the integration over a variable $t$ with a standard normal distribution:
\begin{align}
 \DD t = \dd t \, \NN(t;0,1).
\end{align}
For any function $h$ and integer $i$, we define the $i$-th moment of the product of $h$ multiplied by a Gaussian:
\begin{equation}
 f_i^h(\xh,\xb) = \int \dd x \, x^i h(x) \NN(x;\xh,\xb)	.		\label{eq:f}
\end{equation}
With~(\ref{eq:f}), we define the mean and the variance of the distribution $\frac{h(x)\NN(x;\xh,\xb)}{\f_0^h(\xh,\xb)}$:
\begin{align}
 \fh^h(\xh,\xb) &= \frac{\f_1^h(\xh,\xb)}{\f_0^h(\xh,\xb)} , 	\label{eq:fh}  \\
 \fb^h(\xh,\xb) &= \frac{\f_2^h(\xh,\xb)}{\f_0^h(\xh,\xb)} - \fh^h(\xh,\xb)^2 ,	\label{eq:fb}
\end{align}
It can be verified that following relations hold:
\begin{align}
 \frac{\partial}{\partial \xh} \f_i^h(\xh,\xb) &= \frac{1}{\xb} \left( \f_{i+1}^h(\xh,\xb) - \xh \f_i^h(\xh,\xb) \right) ,	\label{eq:fpartial1} \\
 \frac{\partial}{\partial \xb} \f_i^h(\xh,\xb) &= \frac{1}{2 \xb^2} \left( \f_{i+2}^h(\xh,\xb) -2 \xh \f_{i+1}^h(\xh,\xb) - (\xb - \xh^2)\f_i^h(\xh,\xb) \right) ,	\label{eq:fpartial2} \\
 \frac{\partial}{\partial s }f^h_i(\sqrt{s}t,\rho -s)   &= -\frac{e^{\frac{t^2}{2}}}{2 s}  \frac{\partial}{\partial t} \left( e^{-\frac{t^2}{2}} \frac{\partial}{\partial t} f^h_{i}(\sqrt{s}t,\rho -s) \right) . \label{eq:deOuf} 	
\end{align}
Finally, we introduce two further useful auxiliary functions:
\begin{align}
 \gh^h(\xh,\xb) &= \frac{ \fh^h(\xh,\xb) - \xh}{\xb},   & \gb^h(\xh,\xb) &= \frac{\fb^h(\xh,\xb) - \xb}{\xb^2}. \label{eq:gh}
\end{align}

\section{Algorithms}

\subsection{Message-passing algorithm}
\label{section:bp}
In this paper, we will focus on an approximate message passing (AMP)
algorithm. AMP algorithms originated in studies of problems related to linear estimation \cite{kabashima2003cdma,donoho2009message,gamp}.
For the above probabilistic model of matrix compressed sensing, 
AMP was derived and called \pbig \, in~\cite{parker2015parametric}.
In the following, we explain its principle and expose the main steps of its derivation.

In Bayesian inference, one seeks to produce estimators $\Uhv$ and  $\Vhv$ of $\Uv$ and $\Vv$ 
using the following posterior probability:
\begin{align}
\p(\Uv,\Vv|\y,\A) &\propto \p_U(\Uv) \p_V(\Vv) \p_{Y|Z}\left(\y | \A(\Uv \Vv^{\top}) \right).  \label{eq:proba}
\end{align}
As explained above, the probability distributions used in~(\ref{eq:proba}) ideally match 
the distributions~(\ref{eq:outputChannel}, \ref{eq:priors0}) used for the generation of the problem, 
in which case the inference is said to be Bayes-optimal.
However, it is often the case that these distributions are not known exactly: in this case, the distributions used 
in~(\ref{eq:proba}) are assumptions that we make on the signals' distributions and on the measurement channel.
Inference is in that case suboptimal. 
However, in similar problems it has turned out that the results can still be 
satisfying despite the mismatch between the priors and the actual probability distributions. 
Furthermore, it is possible to parametrize the priors and learn the parameters during inference, 
for example using an expectation maximization procedure~\cite{EM}, which has proven to give satisfying results~\cite{krzakaCS}.

Starting from the posterior probability distribution~(\ref{eq:proba}), the two interesting questions are how to 
evaluate this quantity and how to obtain estimators $(\Uhv,\Vhv)$ from it.
For the second point, we will use the minimal mean squared error (MMSE) estimator, as our goal is to 
obtain low MSEs for~(\ref{eq:mses}). 
Concerning the first point, the problem in estimating~(\ref{eq:proba}) is that it is a distribution in a 
high-dimensional space. Though it is possible to sample from such a distribution using a Monte Carlo Markov chain,
the procedure is very time consuming. Therefore we resort to loopy belief propagation (BP) to estimate the marginals of~(\ref{eq:proba}).
Though not guaranteed to converge on this type of problems, BP has proven to be very successful in a variety of 
similar inference problems~\cite{mezardMontanari,krzakaCS}.

\begin{figure}
 \centering
 \includegraphics[width=0.5\columnwidth]{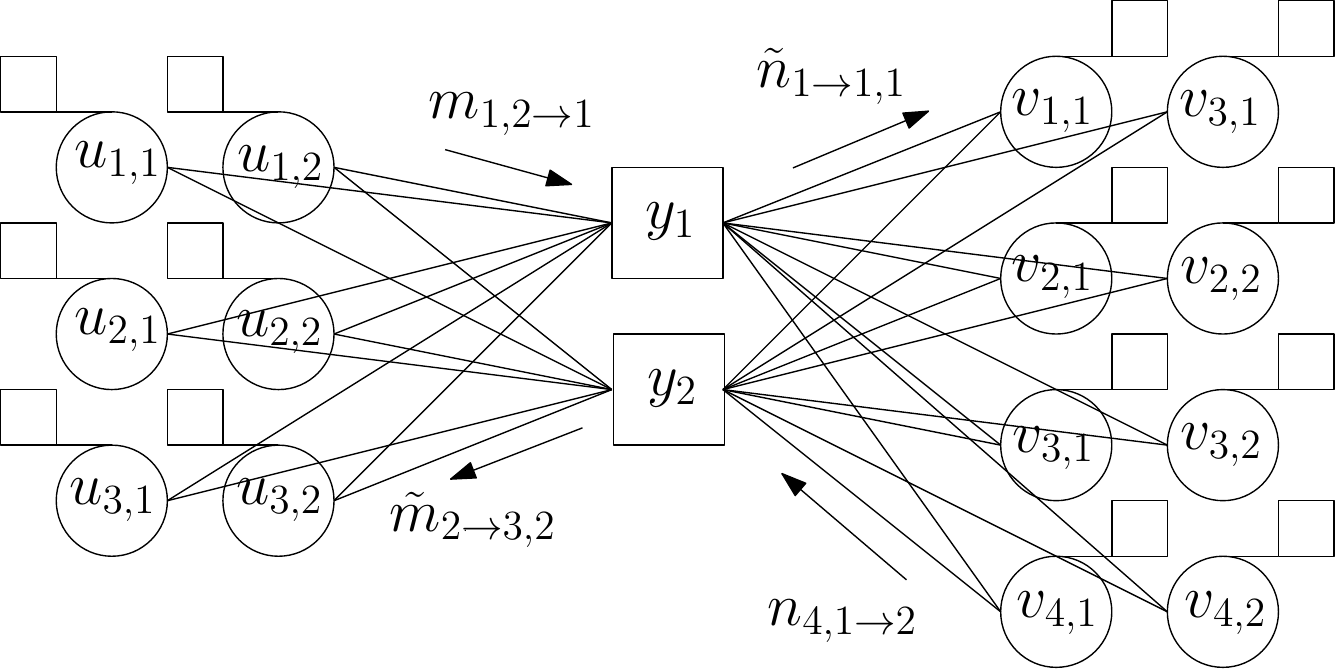}
 \caption{Factor graph associated to the probability
   distribution~(\ref{eq:proba_expanded}). 
Here, we used $\rank=2$, $\US=3$, $\VS=4$, $\YS=2$. Circle represent variables, squares represent constraints.
 The small squares represent the priors on the variables $u$ and $v$. Messages $(\messu, \messuh, \messv, \messvh)$ are sent along each edge of the factor graph.}
 \label{fig:factorGraph}
\end{figure}

In order to derive the BP algorithm, we first rewrite~(\ref{eq:proba}) to make  all variables appear individually:
\begin{align}
 \p(\Uv,\Vv,|\y,\A) \propto \prod_{\ui \ri} &\p_U(u_{\ui \ri})  \prod_{\vi \ri} \p_V(v_{\vi \ri})  
 \int \prod_{\yi} \dd z_{\yi} \p_{Y|Z}(y_{\yi}|z_{\yi}) \delta \left(z_{\yi} - \sum_{\vi=1}^{\VS} \sum_{\ui=1}^{\US} \as^{\vi \ui}_{\yi} \sum_{\ri=1}^{\rank} u_{ \vi \ri} v_{\ui \ri} \right). \label{eq:proba_expanded} 
\end{align}
This probability distribution can be represented by the factor graph in figure~\ref{fig:factorGraph}. 
On it, two types of message pairs $(\messu,\messuh)$ and $(\messv,\messvh)$ are sent to and from the $u$ and $v$ variables respectively.
As the roles of $u$ and $v$ are completely symmetric, we will only treat explicitly the pair $(\messu,\messuh)$: the result can be generalized straightforwardly to $(\messv,\messvh)$.
The message-passing update equations read:
\begin{align}
 \messu_{\ui \ri \to \yi}^{t+1}(u_{\ui \ri}) &\propto \p_U(u_{\ui \ri}) \prod_{\yibis \neq \yi} \messuh_{\yibis \to \ui \ri}^{t}(u_{\ui \ri})   \label{eq:mess_u} , \\
 \messuh_{\yi \to \ui \ri}^{t+1}(u_{\ui \ri}) &\propto \int \left( \prod_{\vi \ribis} \dd v_{\vi \ribis} n_{\vi \ribis \to \yi}^{t+1}(v_{\vi \ribis}) \prod_{(\ribis, \uibis) \neq (\ri, \ui)} \dd u_{\ui \ribis} \messu_{\ui \ribis \to \yi}^{t+1}(u_{\ui \ribis}) \right) 
 \dd z \p_{Y|Z}(y_{\yi}|z) \delta(z - \A(\Uv \Vv^{\top})) ,  \label{eq:mess_u_h}
\end{align}
where the $\propto$ sign stands because $(\messu, \messuh)$ are probability distributions and must therefore be normalized.
These equations can be seen as fixed point equations or as iterative equations that constitute an algorithm. 
For notational lightness, we will do the following calculations without time indices.
However, the correct time indices are crucial for the final algorithm to converge.

A first simplification can be made by replacing the $\rank(\US+\VS)$ integrals in~(\ref{eq:mess_u_h}) 
by a single one over the variable  $z$, which is the sum of $\rank(\US+\VS) -1$ random variables. 
In BP, we assume these random variables to be independent, which allows us to use the central limit theorem.
Calling $\uh_{\ui \ri \to \yi}$ and $\ub_{\ui \ri \to \yi}$ respectively the means and variances of
the variable $u_{\ui \ri}$ distributed according to the distribution $\messu_{\ui \ri \to \yi}$ (and similarly 
for the variables $v_{\vi \ri}$), the variable $z_{\yi} = \sum_{\ui \vi} \as_{\yi}^{\ui \vi} \sum_{\ri} u_{\ui \ri} v_{\vi \ri}$ is a Gaussian variable with mean and variance:
\begin{align}
 \Zh_{\yi} &= \sum_{\ui \vi \ri} \as_{\yi}^{\ui \vi} \uh_{\ui \ri \to \yi} \vh_{\vi \ri \to \yi} ,  \label{eq:zh} \\
 \Zb_{\yi} &= \sum_{\ui \vi \ri} (\as_{\yi}^{\ui \vi})^2 \left[  \ub_{\ui \ri \to \yi} \vb_{\vi \ri \to \yi} + (\uh_{\ui \ri \to \yi})^2 \vb_{\vi \ri \to \yi}  + \ub_{\ui \ri \to \yi}(\vh_{\vi \ri \to \yi})^2 \right] \nonumber   \\
		&+\sum_{\vi \ri} \sum_{\ui \neq \uibis} \as_{\yi}^{\ui \vi} \as_{\yi}^{\uibis \vi} \vb_{\vi \ri \to \yi} \uh_{\ui \ri \to \yi} \uh_{\ui \ribis \to \yi}	    \nonumber \\
		&+\sum_{\ui \ri} \sum_{\vi \neq \vibis} \as_{\yi}^{\ui \vi} \as_{\yi}^{\ui \vibis} \ub_{\ui \ri \to \yi} \vh_{\vi \ri \to \yi} \vh_{\vi \ribis \to \yi} .	      \label{eq:zb}
\end{align}
However, in eq.~(\ref{eq:mess_u_h}), $u_{\ui \ri}$ is fixed and thus $(\uh_{\ui \ri \to \yi}, \ub_{\ui \ri \to \yi})$ has to be replaced by $(u_{\ui \ri},0)$
in~(\ref{eq:zh},\ref{eq:zb}). 
Defining $(\Zh_{\yi \to \ui \ri}, \Zb_{\yi \to \ui \ri})$ to be $(\Zh_{\yi},\Zb_{\yi})$ with $(\uh_{\ui \ri \to \yi}, \ub_{\ui \ri \to \yi}) = (0,0)$ and
\begin{align}
 F_{\yi \ui \ri} &= \sum_{\vi} \as_{\yi}^{\ui \vi} \vh_{\vi \ri \to \yi} , \\
 H_{\yi \ui \ri} &=2 \sum_{\vi} \sum_{\uibis \neq \ui} \as_{\yi}^{\ui \vi} \as_{\yi}^{\uibis \vi} \uh_{\ui \ribis \to \yi} \vb_{\vi \ri \to \yi} ,\\
 G_{\yi \ui \ri} &= \sum_{\vi} (\as_{\yi}^{\ui \vi})^2 \vb_{\vi \ri \to \yi}  ,  
\end{align}
one can rewrite~(\ref{eq:mess_u_h}) with a single integral over a variable $z$ 
following a Gaussian distribution. 
Using the definition~(\ref{eq:f}), the message~(\ref{eq:mess_u_h}) can be expressed as a simple function of the mean and variance of this Gaussian:
\begin{align}
 \messuh_{\yi \to \ui \ri}(u_{\ui \ri}) \propto f_0^Y \left( \Zh_{\yi \to \ui \ri} + F_{\yi \ui \ri} u_{\ui \ri} , \Zb_{\yi \to \ui \ri} + H_{\yi \ui \ri} u_{\ui \ri} + G_{\yi \ui \ri} u_{\ui \ri}^2 \right) .  \label{eq:messuhNice}
\end{align}
Here, we use the simplified notation $f_i^{Y} \equiv f_i^{\p_{Y|Z}}$.
In appendix~\ref{app:amp}, we show how by making a Taylor expansion of this equation, 
we can express the message~(\ref{eq:mess_u}) as
\begin{align}
 \messu_{\ui \ri \to \yi}(u_{\ui \ri}) &\propto p(u_{\ui \ri}) \NN\left( u_{\ui \ri}; \Uh_{\ui \ri \to \yi}, \Ub_{\ui \ri \to \yi} \right),  \label{eq:messuNice}
\end{align}
with
\begin{align}
\Ub_{\ui \ri \to \yi} &=   -\left( \sum_{\yibis \neq \yi}  \left( F_{\yibis \ui \ri}^2 + G_{\yibis \ui \ri} \right) \bar{g}_{\yibis \to \ui \ri} + G_{\yibis \ui \ri} \gh_{\yibis \to \ui \ri}^2  \right)^{-1},  \label{eq:Ub_final}  \\
\Uh_{\ui \ri \to \yi} &=  \Ub_{\ui \ri \to \yi} \sum_{\yibis \neq \yi} F_{\yibis \ui \ri}  \gh_{\yibis \to \ui \ri}, \label{eq:Uh_final}					
\end{align}
where
\begin{align}
 \gh_{\yibis \to \ui \ri} &= \gh^Y(\Zh_{\yibis \to \ui \ri}, \Zb_{\yibis \to \ui \ri}),   &  \gb_{\yibis \to \ui \ri} &= \gb^Y(\Zh_{\yibis \to \ui \ri}, \Zb_{\yibis \to \ui \ri}),
\end{align}
and $(\gh^Y(\cdot,\cdot),\gb^Y(\cdot,\cdot))$ are simplified notations for the functions  $(\gh^{\p_{Y|Z}}(\cdot,\cdot),\gb^{\p_{Y|Z}}(\cdot,\cdot))$  defined in~(\ref{eq:gh}).

This allows us to have a simple expression for the previously introduced mean and variance $\uh_{\ui \ri \to \yi}$ and $\ub_{\ui \ri \to \yi}$ of the message~(\ref{eq:messuNice}).
Using the notations~(\ref{eq:fh}, \ref{eq:fb}), 
\begin{align}
 \uh_{\ui \ri \to \yi} &= \fh^U\left( \Uh_{\ui \ri \to \yi}, \Ub_{\ui \ri \to \yi} \right), & \ub_{\ui \ri \to \yi} &= \fb^U\left(\Uh_{\ui \ri \to \yi}, \Ub_{\ui \ri \to \yi} \right),
\end{align}
where as before, we introduce the simplifying notation $f^U \equiv f^{p_U}$.
As noted previously, the exact same thing can be done for the messages $(\messv,\messvh)$.
The result is an iterative set of equations on a set of means and variances
\begin{align}
 \left( \Zh_{\cdot \to \cdot}^t,\Zb_{\cdot \to \cdot}^t,\gh_{\cdot \to \cdot}^t,\gb_{\cdot \to \cdot}^t,\Uh_{\cdot \to \cdot}^t,\Ub_{\cdot \to \cdot}^t,\uh_{\cdot \to \cdot}^t,\ub_{\cdot \to \cdot}^t,\Vh_{\cdot \to \cdot}^t,\Vb_{\cdot \to \cdot}^t,\vh_{\cdot \to \cdot}^t,\vb_{\cdot \to \cdot}^t     \right)   \label{eq:all_bp_messages}
\end{align}
that constitute the message-passing algorithm.

This algorithm can be further simplified using the so-called Thouless-Andersen-Palmer (TAP) approximation introduced in the study of spin glasses~\cite{TAP}.
We refer the reader to other works in which these simplifications are treated in details~\cite{kabashima2014phase,parker2015parametric} and only give the resulting 
algorithm~\ref{algo:tap}, in which only local quantities and no messages are updated. 
This algorithm is a special case of the ``P-BiG-AMP'' algorithm, introduced
in~\cite{parker2015parametric}.

As its counterparts for generalized linear models (GAMP~\cite{gamp}) or matrix factorization~\cite{kabashima2014phase,bigamp1}, algorithm~\ref{algo:tap} 
needs some adaptations that improve its convergence. 
One very simple damping scheme that allows to improve convergence (though not guaranteeing it) consists in damping a single variable:
\begin{align}
 \Uhv_{t+1} \leftarrow \beta \Uhv_{t+1} + (1-\beta) \Uhv_{t},	\label{eq:simpleDamping}
\end{align}
with $\beta=0.3$, applied right after the calculation of $\Uhv_{t+1}$.
A more involved and better performing, adaptive damping strategy is presented in~\cite{vilaAdaptive}.
Notice that we defined the operators $\Au$ and $\Av$ used in algorithm~\ref{algo:tap} as linear applications $\Au: \RR^{\VS} \to \RR^{\YS \times \US}$ and 
$\Av: \RR^{\US} \to \RR^{\YS \times \VS}$ in~(\ref{eq:Au},\ref{eq:Av}): In the algorithm, we apply them row-wise on the matrices they act on.

\begin{algorithm}[H]
\caption{\pbig \, for matrix compressed sensing}
\label{algo:tap}
  \textbf{Initialization:}  \\
  Initialize the means $(\uhv_0,\vhv_0,\hat{\g}_0)$ and the variances $(\ubv_0,\vbv_0)$ at random according to the distributions $\p_U^0$ and $\p_V^0$.

  \textbf{Main loop:} while $t<t_{\rm max}$, calculate following quantities:
  \begin{align*}
  \bar{\X}_{t+1} &= \ubv_t \vbv_t^{\top} + \ubv_t ( \vhv_t^2 )^{\top} + \uhv_t^2 \vbv_t^{\top} \\
  \hat{\X}_{t+1} &= \uhv_t \vhv_t^{\top} \\
  \Zbv_{t+1} &= \A^2(\bar{\X}_{t+1}) \\
  \Zhv_{t+1} &= \A(\hat{\X}_{t+1}) - \ghv_t \odot \left( \ubv_t \left( \Au(\vhv_t) \odot \Au(\vhv_{t-1}) \right)^{\top} +  \left( \Av(\uhv_t) \odot \Av(\uhv_{t-1}) \right) \vbv_t^{\top} \right) \\
  \gbv_{t+1} &= \gb^Y(\Zhv_{t+1}, \Zbv_{t+1}) \\
  \ghv_{t+1} &= \gh^Y(\Zhv_{t+1}, \Zbv_{t+1}) \\
  \Ubv_{t+1} &= - \left( \left[ \Au(\vhv_t)^2 + \Au^2(\vbv_t) \right] \gbv_{t+1} + \Au^2(\vbv_t) \ghv_{t+1}^2 \right)^{-1} \\
  \Uhv_{t+1} &= \Ubv_{t+1} \odot \left( \Au(\vhv_t) \ghv_{t+1} - \uhv_t \odot \Au(\vhv_t)^2 \gbv_{t+1} - \uhv_{t-1} \odot \Au^2(\vbv_{t-1}) \ghv_{t+1} \odot \ghv_t \right)  \\
  \ubv_{t+1} &= \fb^U(\Uhv_{t+1}, \Ubv_{t+1}) \\
  \uhv_{t+1} &= \fh^U(\Uhv_{t+1}, \Ubv_{t+1})  \\
  \Vbv_{t+1} &= - \left( \left[ \Av(\uhv_t)^2 + \Av^2(\ubv_t) \right] \gbv_{t+1} + \Av^2(\ubv_t) \ghv_{t+1}^2 \right)^{-1} \\
  \Vhv_{t+1} &= \Vbv_{t+1} \odot \left( \Av(\uhv_t) \ghv_{t+1} - \vhv_t \odot \Av(\uhv_t)^2 \gbv_{t+1} - \vhv_{t-1} \odot \Av^2(\ubv_{t-1}) \ghv_{t+1} \odot \ghv_t \right)  \\
  \vbv_{t+1} &= \fb^V(\Vhv_{t+1}, \Vbv_{t+1}) \\
  \vhv_{t+1} &= \fh^V(\Vhv_{t+1}, \Vbv_{t+1})   
  \end{align*}
  \textbf{Result : } $(\Uhv,\Vhv,\hat{\X}, \Zhv)$ are the estimates for $(\Uv,\Vv,\X,\Zv)$ and $(\Ubv,\Vbv,\bar{\X},\Zbv)$ 
  are variances of these estimates.
\end{algorithm}

\section{Asymptotic analysis}
The problem of low-rank matrix compressed sensing can be analyzed with statistical physics methods in the thermodynamic limit,
i.e. when the dimensions of the signals
 $\US$ and $\VS$ and of the measurements $\YS$ go to infinity. $\rank$ can remain finite or go to infinity as well.
 On the other hand, the ratios defined in~(\ref{eq:alphas}) have to be fixed and finite. 
 As in related inference problems, the analysis is done with the replica method.
 The resulting state evolution equations describe 
 the behavior of the corresponding message-passing algorithm.
 In this section, we will focus on the derivation of the replica analysis that results in a simple set of state evolution equations.
 The analysis is very similar to the one of related inference problems~\cite{nishimoriBook,krzakaCS,kabaGAMP,kabashima2014phase}.

\subsection{Replica analysis: free entropy}
Treating an inference problem as a statistical physics problem reduces to writing an energy function corresponding to the problem
and studying the free energy of the system. We are thus interested in calculating a partition function.
Here, the relevant partition function is the normalization constant of the probability distribution~(\ref{eq:proba}):
\begin{equation}
 \ZZ(\yv,\AM) = \int \dd \Uv \, \p_U(\Uv) \int \dd \Vv \, \p_V(\Vv) \int \dd \zv \p_{Y|Z}\left(\yv | \zv \right) \delta \left[ \zv - \A(\Uv \Vv^{\top}) \right]. \label{eq:partition}
\end{equation}
The free entropy $\log \ZZ(\yv,\AM)$ of a given instance can be calculated from the marginals calculated by the belief propagation equations.

However, one can also be interested in the average free entropy of this problem. 
In order to do this, one needs to average $\log \ZZ(\yv,\AM)$ over all possible realizations of $\AM$ and $\yv$, 
for which we use the replica method~\cite{mezardMontanari,nishimoriBook}.
It uses the  identity
\begin{align}
 \langle \log \ZZ \rangle &= \lim_{n \to 0} \frac{\partial}{\partial n} \langle \ZZ^n \rangle     \label{eq:replicaTrick}
\end{align}
where $\langle \cdot \rangle$ denotes the average over $\AM$ and $\y$,
and relies on the fact that an expression for $\ZZ^n$ can be found for integer~$n$. 
This expression is then used for calculating the $n \to 0$ limit in~(\ref{eq:replicaTrick}). Though not rigorous, this method 
has proven to give correct results in a wide range of problems~\cite{mezardMontanari,nishimoriBook}.

Let us therefore start by calculating
\begin{align}
\ZZ(\yv,\AM)^n = \int \prod_{a=1}^n \left\{ \dd \Uv^a \, \p_U(\Uv^a)  \dd \Vv^a \, \p_V(\Vv^a)  \dd \zv^a \p_{Y|Z}\left(\yv|\zv^a \right) \delta\left[ \zv^a - \A(\Uv^a (\Vv^a)^{\top}) \right] \right\} \label{eq:partitionReplicated} 
\end{align}
and its average with respect to the realizations of $\yv$, generated by $\Uv^0$, $\Vv^0$ and $\A$:
\begin{align}
 \langle \ZZ^n \rangle  = \int &\dd \Uv^0 \, \p_U^0(\Uv^0) \dd \Vv^0 \, \p_V^0(\Vv^0) \dd \mathbf{\AM} \, \p_{A}^0( \mathbf{\AM}) \dd \yv  \nonumber \\
    &\dd \zv^0 \p_{Y|Z} (\yv|\zv^0) \delta\left[ \zv^0 - \A(\Uv^0 (\Vv^0)^{\top}) \right] \ZZ(\yv, \AM)^n . \label{eq:partitionReplicatedAveraged}
\end{align}
The indices $a$ represent so-called replicas of the system and are initially independent from each other.
Carrying on the calculation requires to couple them. 
To be more precise, each variable $z_{\yi}^a = [\A(\Uv^a (\Vv^a)^{\top})]_{\yi} $ is the sum of a large number of independent random variables and can therefore be approximated as a Gaussian random variable. 
This was done in section~\ref{section:bp} already and allows again to considerably reduce the number of integrals caused by the averaging over $\AM$. 
However, $z_{\yi}^a$ and $z_{\yi}^b$ are not independent, as they are produced with the same operator $\A$. 
We show in appendix~\ref{app:replica} that $\z_{\yi} \equiv (z_{\yi}^0 \dots z_{\yi}^n)$ is a multivariate random Gaussian variable with mean $0$ and covariance matrix $\Qz \equiv \Qu \odot \Qv$, where the elements of the matrices $\Qu$ and $\Qv$ are given by:

\begin{align}
 Q_u^{ab} &\equiv \frac{1}{\US} \sum_{\ui} u_{\ui}^a u_{\ui}^b  ,  & Q_v^{ab} &\equiv \frac{1}{\VS} \sum_{\vi} v_{\vi}^a v_{\vi}^b . \label{eq:defQ}
\end{align}
As in~(\ref{eq:partitionReplicatedAveraged}), these quantities can be anything, we have to integrate over them, such that
\begin{align}
  \langle \ZZ^n \rangle  &= \int \dd \Qu \left[ \int \prod_a \dd \Uv^a \, \p_U^a(\Uv^a) \prod_{\substack{\ri \\ a \leq b}}  \delta\left( \US Q_u^{ab} - \sum_{\ui} u_{\ui \ri}^a u_{\ui \ri}^b \right) \right]  \nonumber \\ 
  &\int \dd \Qv \left[ \int \prod_a \dd \Vv^a \, \p_V^a(\Vv^a) \prod_{\substack{\ri \\ a \leq b}} \delta\left( \VS Q_v^{ab} - \sum_{\vi} v_{\vi \ri}^a v_{\vi \ri}^b \right) \right] \nonumber \\  
  &\prod_{\yi=1}^{\YS} \left[ \int \dd \z_l \NN(\z_l; 0,\Qz) \int \dd y_l \p_{Y|Z}^0(y_l|z_l^0) \prod_{a=1}^n \p_{Y|Z}(y_l|z_l^a) \right].		\label{eq:ZwithDeltas}
\end{align}
Here, we use the convention that $\p^a_{U}=\p_{U}$ if $a\neq 0$.
We now see that the different replicas are coupled via $\Qu$ and $\Qv$ in the first two lines. 
As we did with $\z_{\yi}$, we now introduce the vector $\mathbf{u}_{\vi \ri} = (u_{\vi \ri}^0 \dots u_{\vi \ri}^n)$ (similarly for $\mathbf{v}_{\ui \ri}$) 
and we use the integral representation of the $\delta$ function, introducing the conjugate variables $\Qhu$ and $\Qhv$ (details in appendix~\ref{app:replica}), which leads to
\begin{align}
  \langle \ZZ^n \rangle  &= \int \dd \Qu \dd \Qhu e^{- \frac{\US \rank}{2} \Tr(\Qu \Qhu)} \left[ \prod_{\ui \ri} \dd \mathbf{u}_{\ui \ri} \p_u(\mathbf{u}_{\ui \ri}) e^{\frac{1}{2} \mathbf{u}_{\ui \ri}^{\top} \Qhu \mathbf{u}_{\ui \ri}} \right] \nonumber \\
  &\int \dd \Qv \dd \Qhv e^{-\frac{\VS \rank}{2} \Tr(\Qv \Qhv)} \left[ \prod_{\vi \ri} \dd \mathbf{v}_{\vi \ri} \p_v(\mathbf{v}_{\vi \ri}) e^{\frac{1}{2} \mathbf{v}_{\vi \ri}^{\top} \Qhv \mathbf{v}_{\vi \ri}} \right] \nonumber \\
  & \prod_{\yi=1}^{\YS} \left[ \int \dd \z_l \NN(\z_l; 0,\Qz) \int \dd y_l \p_{Y|Z}^0(y_l|z_l^0) \prod_{a=1}^n \p_{Y|Z}(y_l|z_l^a) \right].		\label{eq:ZwithQhats}
\end{align}
Finally, we assume the distributions of $u_{\ui \ri}$'s, $v_{\vi
  \ri}$'s and $y_{\yi}$'s are the same for every coordinate. 
Using the notations
\begin{align}
 \p_u(\mathbf{u})&=\p_U^0(u^0)\prod_{a>0}\p_U(u^a), & \p_v(\mathbf{v})&=\p_V^0(v^0)\prod_{a>0}\p_V(v^a), & \p_{y|\mathbf{z}}(y|\mathbf{z})&=\p_{Y|Z}^0(y|z^0) \prod_{a>0} \p_{Y|Z}(y|z^a),
\end{align}
this leads to:
\begin{align}
  \langle \ZZ^n \rangle  &= \int \dd \Qu \dd \Qhu e^{- \frac{\US \rank}{2} \Tr(\Qu \Qhu)} \left[ \dd \mathbf{u} \p_u(\mathbf{u}) e^{\frac{1}{2} \mathbf{u}^{\top} \Qhu \mathbf{u}} \right]^{\rank \US}  \nonumber \\
  &\int \dd \Qv \dd \Qhv e^{-\frac{\VS \rank}{2} \Tr(\Qv \Qhv)} \left[  \dd \mathbf{v} \p_v(\mathbf{v}) e^{\frac{1}{2} \mathbf{v}^{\top} \Qhv \mathbf{v}} \right]^{\rank \VS} \nonumber \\
  & \left[ \int \dd \z \NN(\z; 0,\Qz) \int \dd y \p_{y|\mathbf{z}}(y|\mathbf{z}) \right]^{\YS}.
\end{align}
In the ``thermodynamic'' limit, we take $\US$, $\VS$ and $\YS$ going to infinity with constant ratios.
This motivates us to rewrite the last equation under the form
\begin{align}
  \langle \ZZ^n \rangle  &= \int \dd \Qu \Qhu \Qv \Qhv e^{-\US \rank \left[ \ac_n(\Qu,\Qhu,\Qv,\Qhv) \right] }
\end{align}
and to use the saddle point method, according to which
\begin{align}
  \log \left( \langle \ZZ^n \rangle  \right) &= -\US \rank \min_{\Qu,\Qhu,\Qv,\Qhv} \ac_n(\Qu,\Qhu,\Qv,\Qhv).  \label{eq:saddlePoint}
\end{align}
We are therefore left with a minimization problem over the space of the matrices $\Qu,\Qhu,\Qv$ and $\Qhv$, representing $ 2 (n+1) (n+2)$ parameters (as the matrices are symmetric).

\subsection{Replica symmetric assumption}
The idea of the replica symmetric assumption is that the $n$ replicas introduced in (\ref{eq:partitionReplicated}) are all equivalent, as they are purely a mathematical manipulation.
Based on this, we make the assumption that a sensible matrix $\Qu$ does not make any distinction between the $n$ introduced replicas. 
We therefore parametrize $\Qu$ and $\Qhu$ in the following way:
\begin{align}
 \Qu &= \left( \begin{tabular}{>{$}c<{$} | >{$}c<{$} >{$}c<{$} >{$}c<{$}}
               Q_u^0 & m_u & \cdots & m_u \\
               \hline
               m_u &  Q_u &  \cdots & q_u \\
               \vdots & \vdots & \ddots &\vdots \\
               m_u & q_u & \cdots & Q_u
              \end{tabular}
              \right)             
& \Qhu &= \left( \begin{tabular}{>{$}c<{$} |  >{$}c<{$} >{$}c<{$} >{$}c<{$}}
               \hat{Q}_u^0 & \hat{m}_u & \cdots & \hat{m}_u \\
               \hline
               \hat{m}_u &  \hat{Q}_u &  \cdots & \hat{q}_u \\
               \vdots & \vdots  & \ddots &\vdots \\
               \hat{m}_u & \hat{q}_u &  \cdots & \hat{Q}_u
              \end{tabular}
              \right)		\label{eq:replicaSymmetry}              
\end{align}
and similarly for $\Qv$,
allowing to be left with $16$ instead of $ 2 (n+1) (n+2)$ parameters over which to perform the extremization~(\ref{eq:saddlePoint}). 
Furthermore, $Q_u^0$ and $Q_v^0$ are in fact known, as they are the second moments of the priors $\p_U^0$ and $\p_V^0$, and therefore we set
\begin{align}
 \hat{Q}_u^0 &= 0 , & \hat{Q}_v^0 &= 0 ,
\end{align}
and thus the extremization is only over 12 variables: $(m_u, \hat{m}_u,q_u, \hat{q}_u, Q_u , \hat{Q}_u)$ and $(m_v, \hat{m}_v,q_v, \hat{q}_v, Q_v , \hat{Q}_v)$ .

Let us now look in more details at the function $\ac_n$ to extremize:
\begin{align}
 \ac_n(\Qu,\Qv,\Qhu,\Qhv) \equiv &\left[ \frac{1}{2} \Tr \Qu \Qhu - \log \left( \int \dd \ua \p_u(\ua) e^{\frac{1}{2} \ua^{\top} \Qhu \ua} \right) \right] \nonumber  \\
		    + \frac{\US}{\VS} &\left[ \frac{1}{2} \Tr \Qv \Qhv  - \log \left( \int \dd \va \p_u(\va) e^{\frac{1}{2} \va^{\top} \Qhv \va} \right) \right] \nonumber \\
		    - \frac{\YS}{\rank \VS} &\log \left( \int \dd \z \NN(\z;0,\Qz) \int \dd y  \p_{y|\mathbf{z}}(y|\z) \right)  . 		\label{eq:action}
\end{align}
Thanks to the parametrization~(\ref{eq:replicaSymmetry}), the different terms have simple expressions. 
The traces can simply be written as
\begin{align}
 \Tr \Qu \Qhu &= 2 n m \hat{m}_u + n Q_u \hat{Q}_u + n(n-1) q_u \hat{q}_u ,
\end{align}
while we can use that
\begin{align}
 \ua^{\top} \Qhu \ua  &= \hat{Q}^0_u (u^0)^2  + (\hat{Q}_u-\hat{q}_u) \sum_{a>0} (u^a)^2 + \hat{q}_u ( \sum_{a>0} u^a )^2 + 2\hat{m}_u u^0 \sum_{a> 0} u^a
\end{align}
and the Gaussian transformation $e^{\lambda \alpha^2} = \int {\rm D}x \, e^{\alpha \sqrt{2 \lambda} x}$ 
in order to rewrite the integral $\int \dd \ua P_u(\ua) e^{\frac{1}{2} \ua^{\top} \Qhu \ua}$ as 
\begin{align}
\iu^n &= \int \DD{t}   \int \dd u^0 \, \p_U^0(u^0)  \left[ \int \dd u \, \p_U(u) e^{\frac{\hat{Q}_u-\hat{q}_u}{2} u^2 + (t \sqrt{\hat{q}_u} + \hat{m}_u u^0) u} \right]^n .
\end{align}
The third line in (\ref{eq:action}) can be simplified as well. The first step consists in writing the coupled Gaussian random variables $z^0 \dots z^n$ 
as a function of $n$ independent, standard Gaussian random variables $x^a$ ($a\in [1,n]$) and one additional standard random variable $t$ that couples them all:
\begin{align}
 z^0 &= \sqrt{Q_z^0 - \frac{m_z^2}{q_z}} \, x^0 + \frac{m_z}{\sqrt{q_z}} \, t  , & z^a &= \sqrt{Q_z - q_z} \,  x^a + \sqrt{q_z} \, t.
\end{align}
Making the change of variables in the integral, we obtain the following expression for $\int \dd \z \NN(\z;0,\Qz) \int \dd y  P_{y|z}(y|\z)$:
 \begin{align}
 \iz^n = \int \dd y \, \int \DD{t} & \left[ \int \DD{x^0} \p_{Y|Z}^0(y|\sqrt{Q_z^0 - \frac{m_z^2}{q_z}} \, x^0 + \frac{m_z}{\sqrt{q_z}} \, t) \right] \left[ \int \DD{x} \p_{Y|Z}(y^0| \sqrt{Q_z - q_z} \,  x + \sqrt{q_z} \, t) \right]^n.
 \end{align}
Looking back at the replica trick~(\ref{eq:replicaTrick}), 
we have to study the quantity $\lim_{n \to 0} \frac{\partial }{\partial n} \ac_n$ and therefore the quantities 
 \begin{align}
 \iu(\Qh) = \lim_{n \to 0} \frac{\partial}{\partial n} \log \iu^n  &= \int \DD{t}  \left[ \int \dd u^0 \, \p_U^0(u^0)  \log \left[ \int \dd u \, \p_U(u) e^{\frac{\hat{Q}-\hat{q}}{2} u^2 + (t \sqrt{\hat{q}} + \hat{m} u^0) u} \right] \right], \label{eq:iu} 
 \end{align}
as well as its equivalent $\iv$ (obtained by replacing all $u$s by $v$s in~(\ref{eq:iu}))  and
\begin{align}
\iz(\Q) = \lim_{n \to 0} \frac{\partial}{\partial n} \log \iz^n  &= \int \dd y \, \int \DD{t} f_0^{Y,0}( \frac{m}{\sqrt{q}} t , Q^0 - \frac{m^2}{q} ) \log \left( f_0^Y(\sqrt{q}t, Q-q) \right),
\end{align} 
where $f_i^{Y,0} \equiv f_i^{\p_{Y|Z}^0}$.
In the end, we obtain the free entropy $\phi$ as an extremum
\begin{align}
 \phi = - \extr  & \left\{ \left( m_u \hat{m}_u + \frac{1}{2} Q_u \hat{Q}_u - \frac{1}{2} q_u \hat{q}_u -\iu(\Qhu) \right)  \right. \nonumber  \\
	      +  \frac{\US}{\VS} & \left.\left( m_v \hat{m}_v + \frac{1}{2} Q_v \hat{Q}_v - \frac{1}{2} q_v \hat{q}_v  -\iv(\Qhv) \right) - \frac{\YS}{\rank \VS} \iz(\Qu \odot \Qv)	\right\} 	\label{eq:phi}
\end{align}
over a set of $12$ variables. Note that the shift from a minimum in~(\ref{eq:saddlePoint}) to an extremum in the equation 
above is a consequence to the hazardous $n\to 0$ limit in the replica method.

\subsubsection{Equivalence to generalized matrix factorization}

It is interesting to notice that if $\YS = \US \VS$ and $\rank = O(\US)$, this free entropy is the same as in generalized matrix factorization~\cite{kabashima2014phase}.
This is not an entirely obvious fact, as the two problems are different and that they are identical only if $\A$ is the identity: in generalized matrix factorization,
$ \Z = \lr$.

In order to perform the theoretical analysis of generalized matrix \textit{factorization} as in~\cite{kabashima2014phase}, it is important to take the limit $\rank \to \infty$. 
In fact, it is this limit that ensures that each entry of $\Z$ is the sum of a large number of random variables, which allows to consider that it has a Gaussian distribution.
This is a condition both in the derivation of the message-passing algorithm and in the replica analysis.
For that reason, generalized matrix factorization with finite $\rank$ leads to different algorithms and theoretical bounds~\cite{tanaka,tibo}.
However, in matrix compressed sensing, the mixing of coefficients with $\A$ ensures that even if $\rank=1$, each element of $\Z$ can be considered to have 
a Gaussian distribution. Thanks to this, both the algorithm and the analysis are the same, independently of $\rank$.
Note that it would be natural to write the free entropy~(\ref{eq:phi}) with no explicit $\rank$-dependence by introducing a global measurement ratio $\alpha \equiv \frac{\YS}{\rank (\US + \VS)}$.  

Let us examine the case in which $\YS = \US \VS$ and $\rank = O(\US)$ and the two problems are strictly equivalent.
What differentiates the generalized matrix compressed sensing from the generalized matrix factorization case is that $\A$ is not the identity.
However, as $\A$'s coefficients are Gaussian \iid, it is with high probability a bijection when $\YS=\US \VS$, and in this 
sense the mixing step does not introduce any further difficulty into the problem compared to matrix factorization.
If $\YS > \US \VS$, matrix compressed sensing is not ``compressive'' and therefore
easier than the corresponding matrix factorization problem, because more 
measurements are available. 
If $\YS < \US \VS$, matrix compressed sensing is ``compressive''.
 
\subsection{State evolution equations}
 In the previous section, we have derived an expression of the free entropy as an extremum of an action function over a set of parameters.
 In this section, we find self-consistent equations that hold at the values of these parameters extremizing the action.
 Furthermore, these self-consistent equations can be iterated in order to numerically obtain the extrema of the action.
 
 In order to find the extremum in~(\ref{eq:phi}), we simply set all the partial derivatives of $\phi$ to $0$.
 The difficult part is finding expressions for the derivatives of the integrals $\iu, \iv$ and $\iz$, which we detail here.
 First we do the calculation for $\mathcal{I}_U$.
 \begin{align}
  \frac{\partial}{\partial \hat{Q}_u} \iu(\Qhu) &= \int \DD t \int \dd u^0 \p_U^0(u^0) \frac{ \int \dd u \, \p_U(u) u^2  e^{\frac{\hat{Q}_u - \hat{q}_u}{2} u^2 + (t\sqrt{\hat{q}_u} + \hat{m}_u u^0)u}}{\int \dd u \, \p_U(u) e^{\frac{\hat{Q}_u - \hat{q}_u}{2} u^2 + (t\sqrt{\hat{q}_u} + \hat{m}_u u^0)u}} , \nonumber \\
 \frac{\partial}{\partial \hat{q}_u} \iu(\Qhu) &= \int \DD t \int \dd u^0 \p_U^0(u^0) \frac{ \int \dd u \, \p_U(u) \left( - \frac{u^2}{2} + \frac{tu}{2 \sqrt{\hat{q}_u}} \right)  e^{\frac{\hat{Q}_u - \hat{q}_u}{2} u^2 + (t\sqrt{\hat{q}_u} + \hat{m}_u u^0)u}}{\int \dd u \, \p_U(u) e^{\frac{\hat{Q}_u - \hat{q}_u}{2} u^2 + (t\sqrt{\hat{q}_u} + \hat{m}_u u^0)u}}, \nonumber \\
  \frac{\partial}{\partial \hat{m}_u} \iu(\Qhu) &= \int \DD t \int \dd u^0  u^0 \p_U^0(u^0) \frac{ \int \dd u \, \p_U(u) \, u \, e^{\frac{\hat{Q}_u - \hat{q}_u}{2} u^2 + (t\sqrt{\hat{q}_u} + \hat{m}_u u^0)u}}{\int \dd u \, \p_U(u) e^{\frac{\hat{Q}_u - \hat{q}_u}{2} u^2 + (t\sqrt{\hat{q}_u} + \hat{m}_u u^0)u}} .
\end{align}
 If we inject these expressions into the extremization equations of $\phi$ with respect to $\hat{Q}_u, \hat{q}_u, \hat{m}_u$ and use the update functions defined in~(\ref{eq:f})-(\ref{eq:fb}), we obtain
 \begin{align}
 m_u &= \int \DD t \int \dd u^0 \, u^0 \p_U^0(u^0) \fh^U \left( \frac{ \sqrt{\hat{q}_u} t + \hat{m}_u u^0 }{\hat{q}_u - \hat{Q}_u} , \frac{1}{\hat{q}_u - \hat{Q}_u}  \right)  ,  \\
 Q_u-q_u &= \frac{1}{\sqrt{\hat{q}_u}} \int \DD t \, t \int \dd u^0 \p_U^0(u^0) \fh^U \left( \frac{ \sqrt{\hat{q}_u} t + \hat{m}_u u^0 }{\hat{q}_u - \hat{Q}_u} , \frac{1}{\hat{q}_u - \hat{Q}_u}  \right)  , \label{eq:Q_DE_gen_inter} \\
 Q_u &= \int \DD t \, \int \dd u^0 \p_U^0(u^0) \left[ \fb^U \left( \frac{ \sqrt{\hat{q}_u} t +  \hat{m}_u u^0 }{\hat{q}_u - \hat{Q}_u} , \frac{1}{\hat{q}_u - \hat{Q}_u}  \right) + \left( \fh^U \left( \frac{ \sqrt{\hat{q}_u} t + \hat{m}_u u^0 }{\hat{q}_u - \hat{Q}_u} , \frac{1}{\hat{q}_u - \hat{Q}_u}  \right) \right)^2 \right].
\end{align}
These equations can be further simplified by using the transformation $t \leftarrow t+\frac{\hat{m}}{\sqrt{\hat{q}}}u^0$ and integrating by part eq~(\ref{eq:Q_DE_gen_inter}):
\begin{align}
 m_u &= \sqrt{\frac{\hat{q}_u }{\hat{m}_u ^2}} \int \dd t \, f_1^{U,0} \left(\frac{\sqrt{\hat{q}_u }}{\hat{m}_u }t , \frac{\hat{q}_u }{\hat{m}_u^2} \right)\fh^U \left( \frac{ \sqrt{\hat{q}_u } t }{\hat{q}_u  - \hat{Q}_u } , \frac{1}{\hat{q}_u  - \hat{Q}_u }  \right) ,   \label{eq:mu_DE_gen} \\
 Q_u-q_u &= \sqrt{\frac{\hat{q}_u }{\hat{m}_u ^2}} \int \dd t \,  f_0^{U,0} \left(\frac{\sqrt{\hat{q}_u }}{\hat{m}_u }t , \frac{\hat{q}_u }{\hat{m}_u^2} \right)  \fb^U \left( \frac{ \sqrt{\hat{q}_u } t }{\hat{q}_u - \hat{Q}_u} , \frac{1}{\hat{q}_u - \hat{Q}_u} \right) ,  \label{eq:Qu_DE_gen}  \\
 q_u &= \sqrt{\frac{\hat{q}_u}{\hat{m}_u^2}} \int \dd t \,  f_0^{U,0} \left(\frac{\sqrt{\hat{q}_u}}{\hat{m}_u}t , \frac{\hat{q}_u}{\hat{m}_u^2} \right)  \left[ \fh^U \left( \frac{ \sqrt{\hat{q}_u} t }{\hat{q}_u - \hat{Q}_u} , \frac{1}{\hat{q}_u - \hat{Q}_u} \right) \right]^2 ,  \label{eq:qu_DE_gen}
\end{align}
and the same equations hold replacing $u$ by $v$.

Let us now come to the derivatives of $\iz$. 
To calculate them, we use the identity~(\ref{eq:deOuf}), taking $s=q$ or $s=\frac{m^2}{q}$.
After an integration by parts, we obtain
\begin{align}
 \frac{\partial}{\partial m} \iz(\Q) &= \frac{1}{m} \int \dd y \int \DD t \frac{ \left[ \frac{\partial}{\partial t} f_0^{Y,0}( \frac{m}{\sqrt{q}} t , Q^0  - \frac{m^2}{q} ) \right] \left[ \frac{\partial}{\partial t} f_0^{Y}( \sqrt{q} t , Q -q ) \right]}{f_0^{Y}( \sqrt{q} t , Q -q )} , \\
 \frac{\partial}{\partial q} \iz(\Q) &= -\frac{1}{2 q} \int \dd y \int \DD t \left[ \frac{\frac{\partial}{\partial t} f_0^{Y}( \sqrt{q} t , Q -q )}{f_0^{Y}( \sqrt{q} t , Q -q )} \right]^2 f_0^{Y,0}( \frac{m}{\sqrt{q}} t , Q^0  - \frac{m^2}{q} ) , \\
 \frac{\partial}{\partial Q} \iz(\Q) &= \int \dd y \int \DD t f_0^{Y,0}( \frac{m}{\sqrt{q}} t , Q^0  - \frac{m^2}{q} ) \left[ \frac{\frac{\partial}{\partial Q} f_0^{Y}( \sqrt{q} t , Q -q )}{f_0^{Y}( \sqrt{q} t , Q -q )} \right].
\end{align}
Injecting these expressions into the extremization equations of $\phi$ with respect to $Q,q,m$, we obtain
\begin{align}
  \hat{m} &= \frac{1}{m}  \int \dd y \int \DD t \frac{ \left[ \frac{\partial}{\partial t} f_0^{Y,0}( \frac{m}{\sqrt{q}} t , Q^0 - \frac{m^2}{q} ) \right] \left[ \frac{\partial}{\partial t} f_0^{Y}( \sqrt{q} t , Q -q ) \right]}{f_0^{Y}( \sqrt{q} t , Q -q )} ,  \label{eq:mh_DE_gen} \\
  \hat{q} &= \frac{1}{q}  \int \dd y \int \DD t \left[ \frac{\frac{\partial}{\partial t} f_0^{Y}( \sqrt{q} t , Q -q )}{f_0^{Y}( \sqrt{q} t , Q -q )} \right]^2 f_0^{Y,0}( \frac{m}{\sqrt{q}} t , Q^0  - \frac{m^2}{q} ) ,  \label{eq:qh_DE_gen}\\
  \hat{Q} &= 2 \int \dd y \int \DD t f_0^{Y,0}( \frac{m}{\sqrt{q}} t , Q^0  - \frac{m^2}{q} ) \left[ \frac{\frac{\partial}{\partial Q} f_0^{Y}( \sqrt{q} t , Q -q )}{f_0^{Y}( \sqrt{q} t , Q -q )} \right] , \label{eq:Qh_DE_gen}
\end{align}
and remembering that $m=m_u m_v, q=q_u q_v$, $Q = Q_u Q_v$ and the definitions~(\ref{eq:alphas}):
\begin{align}
\hat{m}_u &= \au m_v \hat{m},  &  \hat{q}_u &= \au q_v \hat{q} , &  \hat{Q}_u &= \au Q_v \hat{Q} ,  \label{eq:mhu_mh}\\
 \hat{m}_v &= \av m_u \hat{m},  & \hat{q}_v &= \av q_u \hat{q} , &  \hat{Q}_v &= \av Q_u \hat{Q} . \label{eq:mhv_mh}
\end{align}
 The equations~(\ref{eq:mu_DE_gen},\ref{eq:Qu_DE_gen},\ref{eq:qu_DE_gen}) along with their equivalents for $v$, the equations~(\ref{eq:mh_DE_gen},\ref{eq:qh_DE_gen},\ref{eq:Qh_DE_gen}) and~(\ref{eq:mhu_mh},\ref{eq:mhv_mh}) 
 constitute a closed set of equations that hold at the extrema of $\phi$ in equation~(\ref{eq:phi}).
 
 When they are iterated, they constitute the so-called state evolution equations. 
 These can also be obtained by the analysis of the BP algorithm and are known to accurately describe the 
 algorithm's behavior when the replica symmetric hypothesis is indeed correct.
 
 As noted before, if $\YS = \US \VS$, these state evolution equations are identical to the ones in matrix factorization~\cite{kabashima2014phase}.
 Therefore, they reduce to the state evolution of GAMP when $\U$ is known, which corresponds to fixing $m_u=q_u=Q_u=\smu$ in the equations.

\subsection{Bayes-optimal analysis}
Until now, we have not supposed exact knowledge of the true signal distributions and of the true measurement channel. 
When this is the case, the state evolution equations greatly simplify because of the so-called Nishimori conditions~\cite{lenkaReview}.
In our case, these ensure that following equalities hold:
\begin{align}
 Q = Q^0 , \quad \hat{Q} = 0, \quad m = q, \quad \hat{m} &= \hat{q}	\label{eq:nishConds}
\end{align}
both for $u$ and $v$.
Then, we only need to keep track of the variables $(m_u, \hat{m}_u, m_v, \hat{m}_v)$,
and  the state evolution is obtained by choosing initial values for $(m_u^0, m_v^0)$ and iterating for $\iter \geq 0$ the equations
\begin{align}
 \hat{m}^{\iter+1} &= \frac{1}{m_u^{\iter} m_v^{\iter}} \int \dd y \int \DD t  \frac{\left[ \frac{\partial}{\partial t} f^Y_{0}(\sqrt{m_u^{\iter} m_v^{\iter}}t,\smu \smv  -m_u^{\iter} m_v^{\iter})\right]^2}{f^Y_{0}(\sqrt{m_u^{\iter} m_v^{\iter}}t, \smu \smv -m_u^{\iter} m_v^{\iter})} ,   \label{eq:mh_DE_BO} \\
 m_u^{\iter+1} &= \frac{1}{\sqrt{\au m_v^{\iter} \hat{m}^{\iter+1}}} \int \dd t \frac{\left[f^U_1(\frac{t}{\sqrt{\au m_v^{\iter} \hat{m}^{\iter+1}}},\frac{1}{\au m_v^{\iter} \hat{m}^{\iter+1}})\right]^2}{f^U_0(\frac{t}{\sqrt{\au m_v^{\iter} \hat{m}^{\iter+1}}},\frac{1}{\au m_v^{\iter} \hat{m}^{\iter+1}})} , \label{eq:mu_DE_BO} \\
 m_v^{\iter+1} &= \frac{1}{\sqrt{\av m_u^{\iter} \hat{m}^{\iter+1}}} \int \dd t \frac{\left[f^V_1(\frac{t}{\sqrt{\av m_u^{\iter} \hat{m}^{\iter+1}}},\frac{1}{\av m_u^{\iter} \hat{m}^{\iter+1}})\right]^2}{f^V_0(\frac{t}{\sqrt{\av m_u^{\iter} \hat{m}^{\iter+1}}},\frac{1}{\av m_u^{\iter} \hat{m}^{\iter+1}})} , \label{eq:mv_DE_BO} 
\end{align}
until convergence. From $m_u$ and $m_v$, one can simply deduce the 
mean squared errors by the following relations:
\begin{align}
 \mseu &=  \smu - m_u, & \msev &=  \smv - m_v , & \msex &= \smu \smv- m_u m_v.  \label{eq:mses_BO}
\end{align}
The initialization values $(m_u, \hat{m}_u, m_v, \hat{m}_v)$ indicate how close to the solution the algorithm is at initialization. 
In case of a random initialization of the algorithm, the expected initial overlaps $m_u^0$ and $m_v^0$ are of order $1/\US$ and $1/\VS$ respectively,
and they should therefore be set to these values (or less) in the state evolution equations.

Note that state evolution run with matching priors without imposing the Nishimori conditions~(\ref{eq:nishConds}) should in principle 
give the exact same results as the Bayes-optimal state evolution analysis presented above, and thus naturally follow the so-called ``Nishimori line'' defined by~(\ref{eq:nishConds}).
However, as shown in~\cite{caltaConvergence}, the Nishimori line can be unstable: In that case, numerical fluctuations around it will be amplified 
under iterations of state evolution that will thus give a different result than its counterpart with imposed Nishimori conditions.
This instability of the Nishimori line seems to be the reason why algorithm~\ref{algo:tap} as well as others of the same type do not converge without damping of the variables.

\section{Case Study}
In this section, we focus on one specific setting for which the state evolution equations are practical to implement. 
An analysis of their fixed points leads to an understanding of different phases and of the phase transitions between them.

We look at the setting in which both $\Uv$ and $\Vv$ follow a Bernoulli-Gauss distribution:
\begin{align}
 \p_U(u) &= (1-\rho_u) \delta(u) + \rho_u \, \NN(u;0,1), \\
 \p_V(v) &= (1-\rho_v) \delta(v) + \rho_v \, \NN(v;0,1),
\end{align}
and the measurements are taken through an additive white Gaussian noise (AWGN) channel:
\begin{align}
 \forall \yi \in [1, \YS], \quad Y_{\yi} = [ \A(\Uv \Vv^{T})]_{\yi} + \xi_{\yi}, \qquad \quad \text{with} \quad \xi_{\yi} \sim \NN(\xi_{\yi};0,\Delta).
\end{align}
Note that most previous works~\cite{lee2013near,nuclearNormMin,JainLowRank,laffertyLowRank} consider this channel.
For the AWGN channel the equation~(\ref{eq:mh_DE_BO}) has a simple analytical expression:
\begin{align}
 \hat{m}^{\iter+1} &= \frac{1}{\Delta + \rho_u \rho_v - m_u^{\iter} m_v^{\iter}}.
\end{align}
Further simplifying the setting to the special case $\US=\VS$ and $\rho_u=\rho_v=\rho$, the Bayes optimal state evolution equations~(\ref{eq:mh_DE_BO}-\ref{eq:mv_DE_BO}) can be written as one single equation
\begin{align}
 m &= \sqrt{\frac{\Delta+\rho^2-m^2}{\alpha_u m}} \int \dd t \frac{\left[ f_1^U(\sqrt{\frac{\Delta+\rho^2-m^2}{\alpha_u m}}t , \frac{\Delta+\rho^2-m^2}{\alpha_u m}) \right]^2}{f_0(\sqrt{\frac{\Delta+\rho^2-m^2}{\alpha_u m}}t , \frac{\Delta+\rho^2-m^2}{\alpha_u m})}, \label{eq:simple_DE}
\end{align}
in which the iteration-time indices of $m$, $\iter$ (left hand side) and $\iter-1$ (right hand side), are left out for better legibility. 
We can define a global measurement rate
\begin{align}
 \alpha \equiv \frac{\YS}{2 \US \rank} = \frac{\alpha_u}{2} ,	\label{eq:global_alpha}
\end{align}
which is the natural quantity to compare $\rho$ to.

\subsection{Phases and phase transitions}
As in compressed sensing or in matrix factorization, the analysis of the free entropy and state evolution equations 
reveals the existence of different phases in which the difficulty of the problem is different.
In our case study, the free entropy $\phi$ has the following expression:
\begin{align}
 \phi(m) &= - m \hat{m} - \frac{\alpha}{4} \log\left( 2 \pi \left( \Delta + \rho^2 - m^2\right)\right) \nonumber \\
 &+ \frac{2}{\sqrt{\hat{m}}} \int \dd t f_0^U \left( \frac{t}{\sqrt{\hat{m}}}, \frac{1}{\hat{m}} \right) \left[ \frac{t^2}{2} + \log\left( \sqrt{\frac{2 \pi}{\hat{m}}} f_0^U \left( \frac{t}{\sqrt{\hat{m}}}, \frac{1}{\hat{m}} \right) \right) \right]
\end{align}
with
\begin{align}
  \hat{m} &= \frac{1}{\Delta + \rho^2 - m^2}.
\end{align}
The integral can best be numerically evaluated replacing $\int$ by $2 \left( \int_{0}^{20} + \int_{20}^{20 \sqrt{1+\hat{m}}} \right) $, which allows a reliable 
numerical evaluation for all possible values of $\hat{m}$.

\begin{figure}[h]
 \centering
  \includegraphics[width=0.5\textwidth]{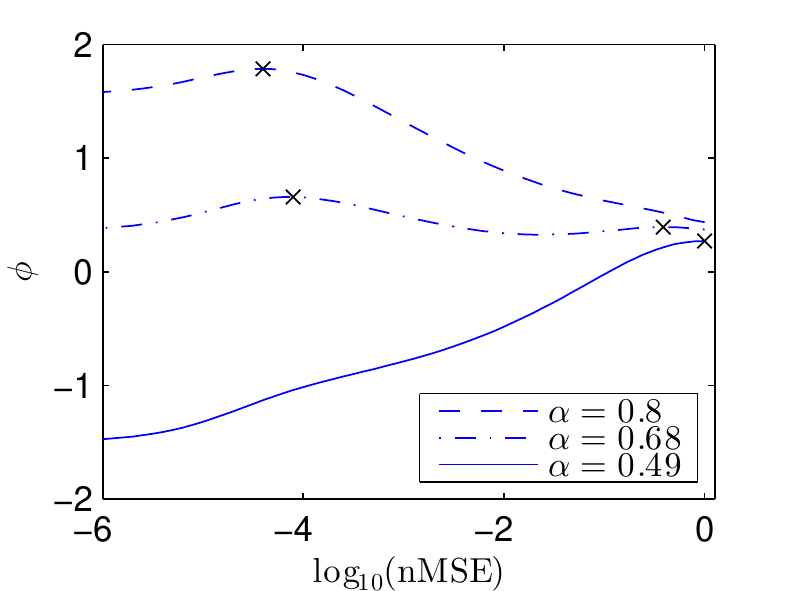}
  \caption{Free entropy landscapes for $\rho=0.5$, $\Delta=10^{-5}$. 
  Crosses represent local maxima. There are three types of them: either at nMSE$=1$ (as for $\alpha=0.49$),
  or at ${\rm nMSE} \approx \Delta$, or in an intermediary region. In case there are several local maxima (as for $\alpha=0.68$), 
  the algorithm will perform sub-optimally, getting stuck in the local maximum of highest nMSE instead of converging to the global maximum (``hard but possible'' phase). }
\label{fig:phases}
\end{figure}

Figure~\ref{fig:phases} shows the free entropy landscapes for $\rho=0.1$ and different values of $\alpha$.
Instead of using $m$ as $x$-axes, we use the normalized mean squared error 
\begin{align}
 {\rm nMSE} = 1 - \frac{m}{\rho}, \label{eq:nMSE}
\end{align}
that is a more natural quantity to measure the quality of reconstruction.

We can define three different phases depending on the positions of the free entropy maxima.
In the noiseless setting, these are:
\begin{enumerate}
 \item An ``impossible'' phase, in which the global maximum of the
   free entropy is not at nMSE$=0$.
 In that phase, no algorithm can find the correct solution.
 \item A ``hard but possible'' phase, in which the free entropy has its global maximum at nMSE$=0$, but also a \textit{local} maximum at non-zero nMSE.
 In that phase, it is possible to find the correct solution, by correctly sampling from the posterior distribution~(\ref{eq:proba}).
 However, algorithms such as \pbig \, get stuck in the local free entropy maximum instead of finding the global maximum.
 \item An ``easy'' phase, in which the free entropy function has a single maximum at nMSE$=0$.
\end{enumerate}
In a noisy setting as in figure~\ref{fig:phases}, the lowest achievable nMSE is of the order of the AWGN variance $\Delta$ instead of $0$.

\subsubsection{State evolution fixed points}
\begin{figure}[h]
 \centering
 \subfloat[$\rho=0.1$]{\label{fig:fp01}%
  \includegraphics[width=0.48\textwidth]{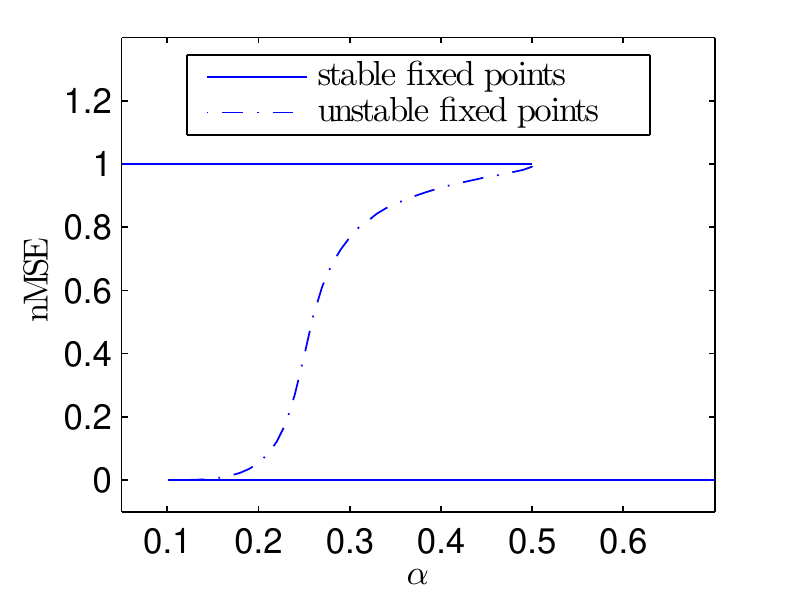}%
}
 \subfloat[$\rho=0.6$]{\label{fig:fp06}%
  \includegraphics[width=0.48\textwidth]{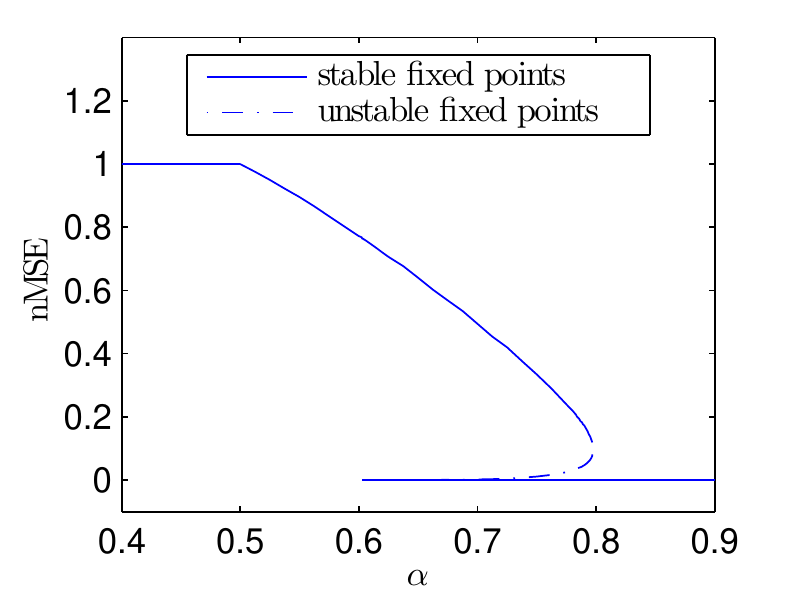}%
}
\caption{Fixed points of the state evolution equation~(\ref{eq:simple_DE}) for two different sparsities $\rho$.
For values of $\alpha$ for which two stable fixed points exist, the iterated state evolution equation converges to the one of higher nMSE 
if the initial nMSE is higher than the unstable fixed point, and to the one of lower nMSE if not.}
\label{fig:fp}
\end{figure}

The state evolution equation~(\ref{eq:simple_DE}) can either be iterated or considered as a fixed point equation.
Figure~\ref{fig:fp} shows the fixed points of~(\ref{eq:simple_DE}),
which are all local extrema of the free entropy $\phi$. 
The iterated state evolution equation converges to one of the local
maxima. Since the
state evolution for the matrix compressed sensing problem and the
dictionary learning problem are the same (provided $L=MP$ and
$R=O(M)$) these diagrams and their analysis are equivalent to those
presented in previous work on the dictionary learning
\cite{kabashima2014phase}. Notably \cite{kabaSample} presented
analogous diagrams depicting the fixed points for the dictionary
learning problem. 

The plots allow to see more clearly the ``impossible'', ``hard but possible'' and ``easy'' phases.
In the ``hard but possible'' phase, the state evolution has an unstable fixed point, which corresponds 
to a local minimum of the free entropy.
Three interesting facts can be noticed:
\begin{enumerate}
 \item In the noiseless setting, the impossible/possible phase
   transition (the apparition of the low ${\rm nMSE}$ fixed point) takes place at $\alpha=\rho$.
 This can be expected because because it is the critical $\alpha$ at which the number of available equations is equal to the total number of non-zero components of the unknowns, just as in compressed sensing.
 \item The fixed point at nMSE=1 always exists and is stable for $\alpha \in [0, 1/2]$. 
 This is a rather remarkable fact that does not appear in compressed
 sensing. A consequence of this is the existence of a ``hard but
 possible'' phase that even for very small values of $\rho$ extends at
 least up to $\alpha=1/2$. 
 This radically differs from the low-$\rho$ regime in compressed
 sensing, in which the measurement rate $\alpha$ necessary for
 tractable recovery goes to zero as $\rho \to 0$.  
 \item Increasing $\alpha$ starting below $1/2$ and following the high-nMSE branch, two successive phase transitions are encountered. 
 First, the ${\rm nMSE}=1$ fixed point disappears at $\alpha=1/2$ and turns into an
 ${\rm nMSE}<1$ fixed point in a second order (i.e. continuous) phase transition.
 Second, the upper branch disappears and the discontinuity of the nMSE of the fixed point, jumping down to the lower branch, marks a first order phase transition.
 While these two transitions of different types are clearly visible in Figure~\ref{fig:fp06}, they are too close together in Figure~\ref{fig:fp01} to be distinguished. 
 They are separated nonetheless, the easy/hard (first order) phase transition always takes place at $\alpha>1/2$.
\end{enumerate}

\begin{figure}[h]
 \centering
 \includegraphics[width=0.5\textwidth]{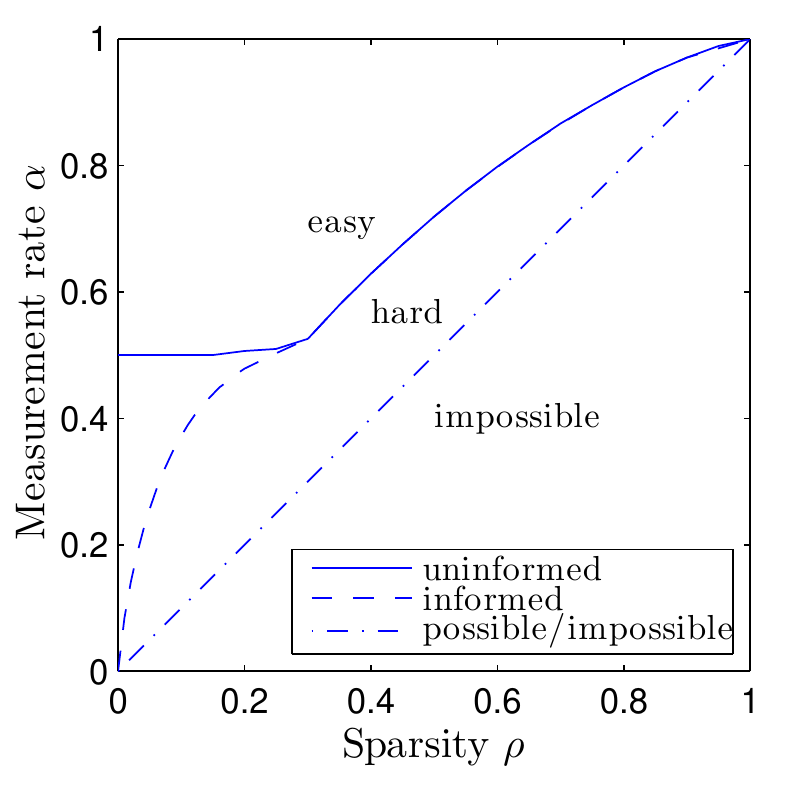}
 \caption{Phase diagram for the considered case-study obtained from
   the state evolution, eq.~(\ref{eq:simple_DE}). Noise variance is  $\Delta=10^{-12}$ and success is defined by a final nMSE$<10^{-10}$.
 The disappearing of the state evolution fixed point (or equivalently,
 of a free entropy maximum) with nMSE of order 1 marks the frontier
 between the ``hard'' and the ``easy'' phase (full line). The dashed
 line marks the easy/hard phase boundary when an ``informed''
 initialization is provided (see text). 
 The possible/impossible frontier represented corresponds to the noiseless case.}
 \label{fig:full_phase_diagram}
\end{figure}

Figure~\ref{fig:full_phase_diagram} shows the full phase diagram for
the case-study problem, with the easy, hard and impossible phases.
The ``uninformed'' line is obtained by starting the state evolution
starting from nMSE$= 1 - \epsilon$, with an infinitesimally small $\epsilon$, and defines the transition between the ``easy'' and the ``hard'' phase.
 Interestingly, the entire region with $\alpha<0.5$ is in the hard
 phase, even at low values of $\rho$, due to the existence of the stable
 fixed point at ${\rm nMSE}=1$.
 In the ``hard'' phase, inference is possible provided a good estimation of the signal is already known. 
 The effect of such a partial knowledge can be simulated by running the state evolution equation (\ref{eq:simple_DE}) starting with nMSE$=0.9$, 
 leading to the ``informed'' line, for which $\alpha \to 0$ when $\rho
 \to 0$. The position of this line depends strongly on the starting nMSE.

\subsection{Comparison with algorithmic performances}
\begin{figure}[h]
 \centering
  \subfloat[$\rho=0.1$, $M=50$]{\label{fig:comp01N50}%
  \includegraphics[width=0.48\textwidth]{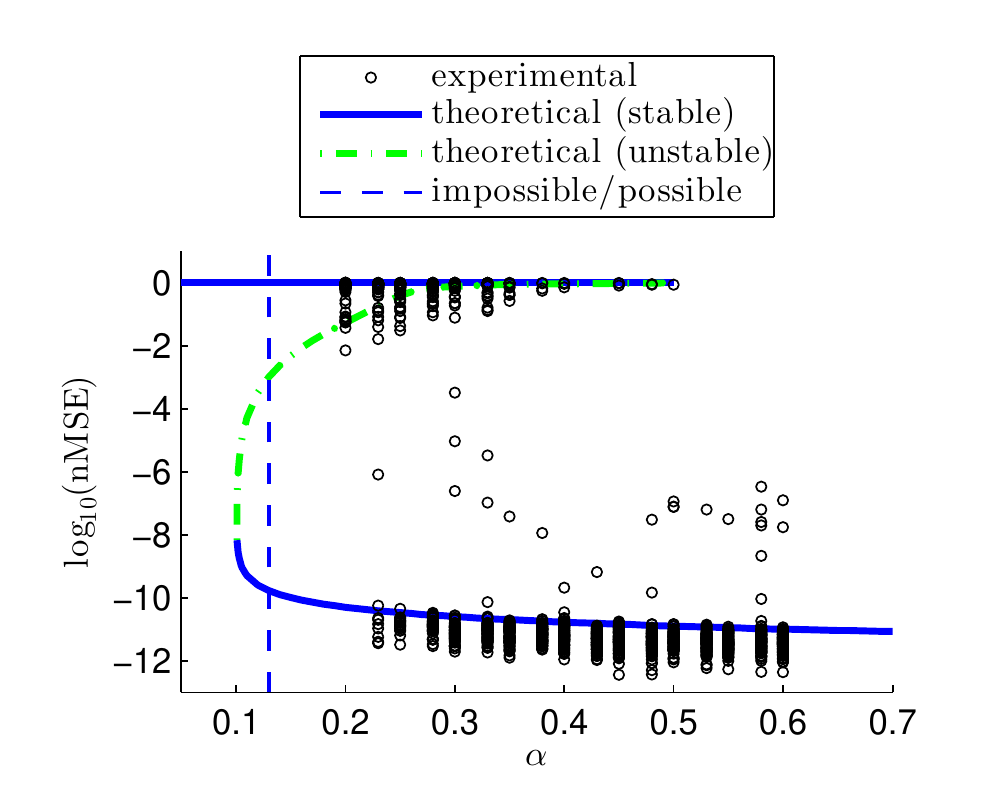}%
}
  \subfloat[$\rho=0.1$,  $M=200$]{\label{fig:comp01N200}%
  \includegraphics[width=0.48\textwidth]{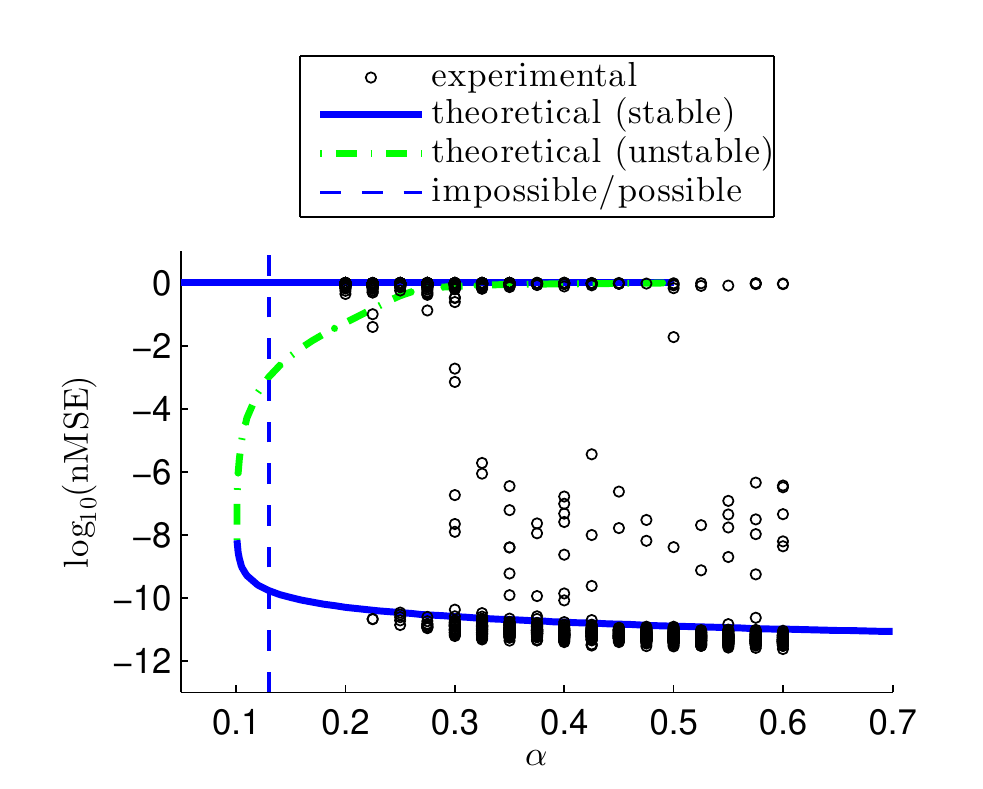}%
}
\caption{Comparison of fixed points obtained by the state evolution
  and and values reached by the \pbig \, algorithm.  
Parameters  are $\rho=0.1$, $\Delta=10^{-12}$ with (a): $M=50$, (b): $M=200$.
For each $\alpha$ there are $100$ experimental points.
The experimental fixed points are relatively close to the fixed points of the state evolution.
Note that the spreading around the theoretical line diminishes with growing $M$.
In the thermodynamic limit $M\to \infty$, all experimental points would be on the fixed point of \textit{highest} nMSE.
At finite $M$, the probability to initialize the algorithm \textit{below} the unstable fixed point allows some instances to 
converge to the low-nMSE fixed point.}
\label{fig:comp01}
\end{figure}

\begin{figure}[h]
 \centering
 \subfloat[$\rho=0.6$, $M=50$]{\label{fig:comp06N50}%
  \includegraphics[width=0.48\textwidth]{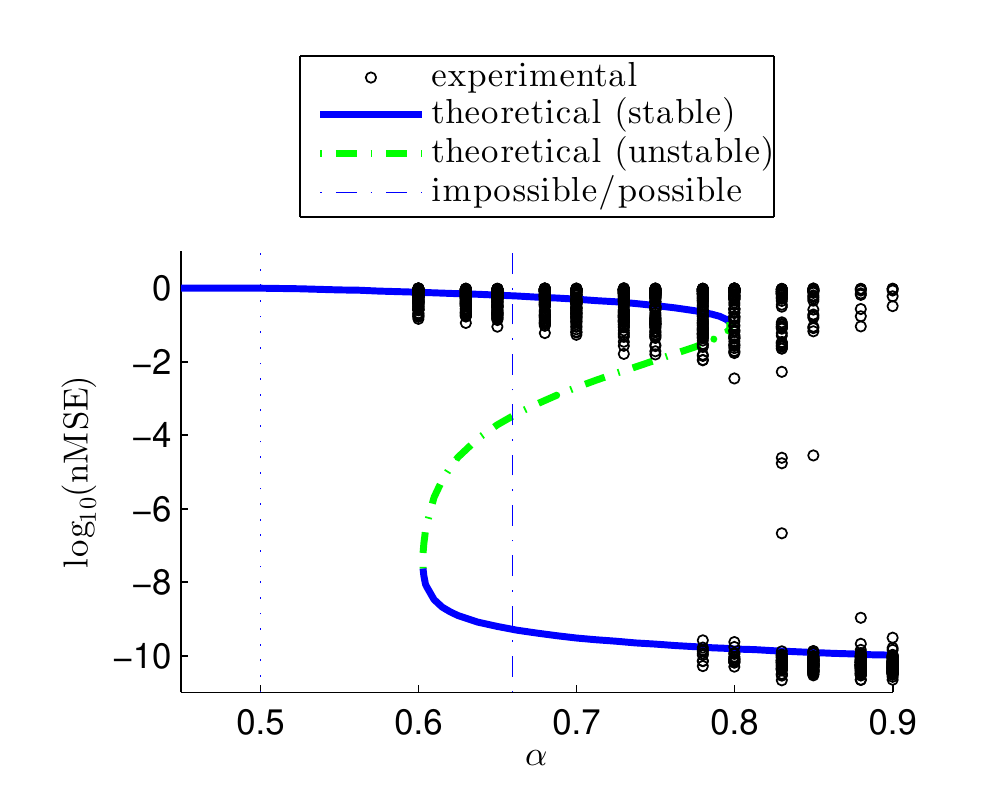}%
}
 \subfloat[$\rho=0.6$,  $M=200$]{\label{fig:comp06N200}%
  \includegraphics[width=0.48\textwidth]{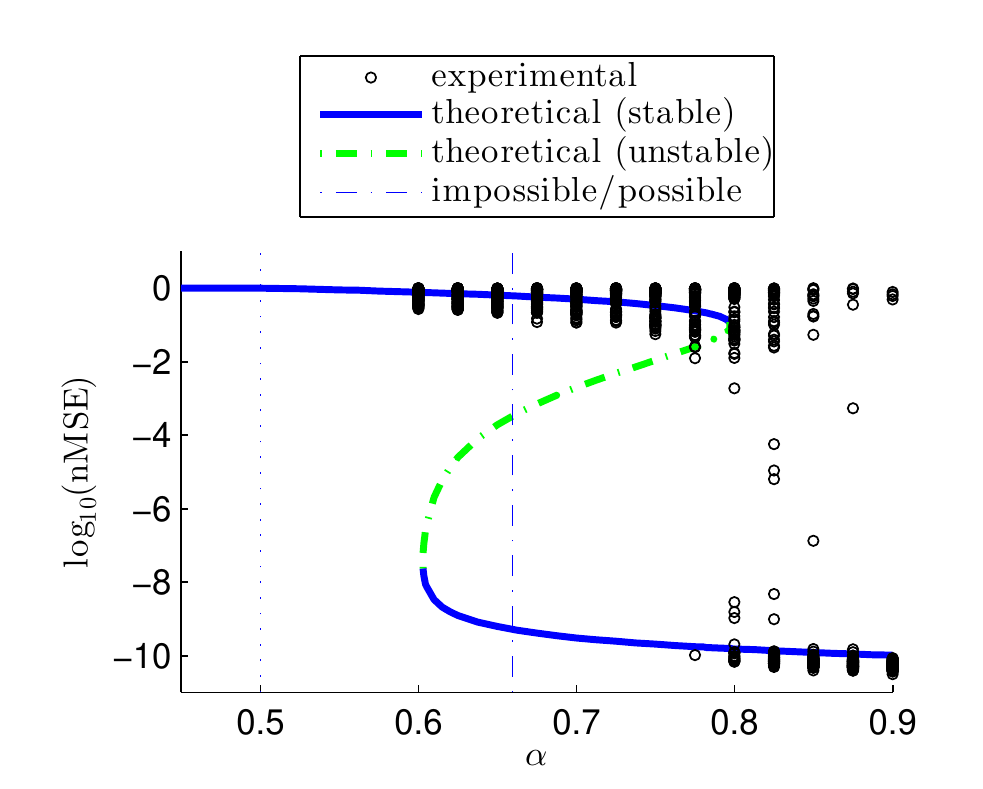}%
}
\caption{Comparison of fixed points obtained by the state evolution
  and values reached by the \pbig \, algorithm.
Parameters  are $\rho=0.6$, $\Delta=3.6\times 10^{-11}$ with (a): $M=50$, (b): $M=200$. 
For each $\alpha$ there are $100$ experimental points.
Unlike for the $\rho=0.1$ case on figure~\ref{fig:comp01},the algorithm 
fails for an important fraction of instances in the ``easy'' phase.
This phenomenon is not explained by the state evolution analysis and 
might be a finite size effect. However, as $\alpha$ grows the probability of success goes to $1$ (see figure~\ref{fig:PT_06}).
Unlike for $\rho=0.1$, the probability of recovery inside the ``hard'' phase is much smaller, due to the lower nMSE of the unstable fixed point.
The thin dotted line marks the position of the second order phase transition, at which the nMSE stops being strictly equal to 1. }
\label{fig:comp06}
\end{figure}

Figures~\ref{fig:comp01} and~\ref{fig:comp06}  presents a comparison of the theoretical fixed point analysis performed above with 
the actual performances of \pbig. 

For the experiments, rank $\rank=1$ was used. In this setting, the only invariance left is a scaling invariance: if $(\Uv,\Vv)$ is the true solution, then for every 
$\gamma \neq 0$, $(\gamma \Uv, \frac{1}{\gamma} \Vv)$ is a solution as well. 
The final nMSE returned by the algorithm takes this invariance into account and is the average of the error on $\Uv$ and the error on $\Vv$:
\begin{align}
\rm{nMSE} = \frac{1}{2} \left( {\rm nMSE}_u + {\rm nMSE}_v \right) 
\end{align}
which will be compared to the results obtained by the theoretical
expression~(\ref{eq:nMSE}). For each instance of the problem, the algorithm was allowed up to $20$ restarts  from different random initializations to reach a nMSE smaller than $10^{-6}$,
and the lowest of the reached nMSE was kept.

The results show that there is a good agreement between the theory and the performance of \pbig: most of the nMSEs reached by \pbig \, correspond to a stable fixed point of the state evolution.
The agreement with the theory becomes better with increasing system size. 
For smaller sizes, the experimental points are more spread around the theoretical fixed points. This can be well understood by analyzing the case of fixed points with nMSE=1.
The ``meaning'' of such fixed points is that the algorithm is unable to estimate the true signals better than at random. 
In the $\US \to \infty$ limit, the nMSE between the true signals and random signals is $1$ with probability $1$.
For finite values of $\US$ however, the nMSE between true and random signals follows a distribution on $[0,1]$ that gets more peaked on $1$ as $\US$ increases.
This explains the narrowing of the spread of experimental points around the fixed points as $\US$ increases.

\subsubsection{Succeeding in the hard phase: importance of the initialization}
An interesting consequence of this finite size effect is that for small $\US$, parts of the ``hard'' phase are quite easy.
The reason is that if the random initialization of the algorithm is such that the nMSE is \textit{smaller} than the nMSE of the \textit{unstable} fixed point, the algorithm naturally converges
 to the low-nMSE solution. Therefore, running the algorithm from a few different initializations can allow to converge to the correct solution even in the ``hard'' phase, 
 provided that $\US$ is small enough and that the unstable fixed point has a high enough nMSE.
 
\begin{figure}[h]
 \centering
  \subfloat[$\rho=0.1$]{\label{fig:PT_01}%
  \includegraphics[width=0.40\textwidth]{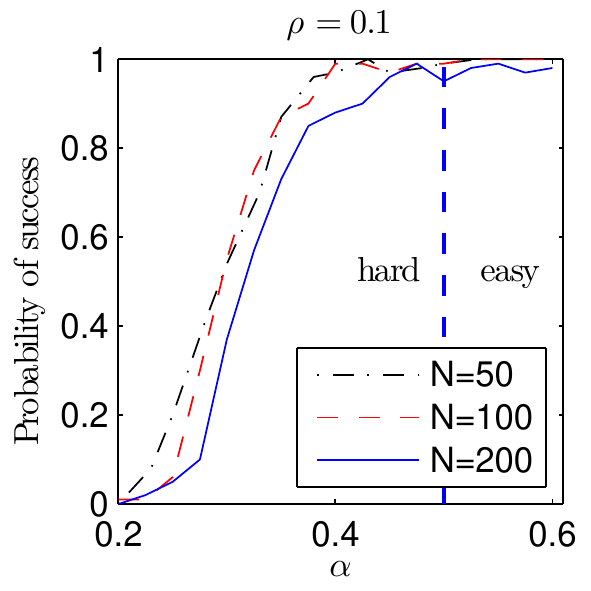}%
}
  \subfloat[$\rho=0.6$]{\label{fig:PT_06}%
  \includegraphics[width=0.40\textwidth]{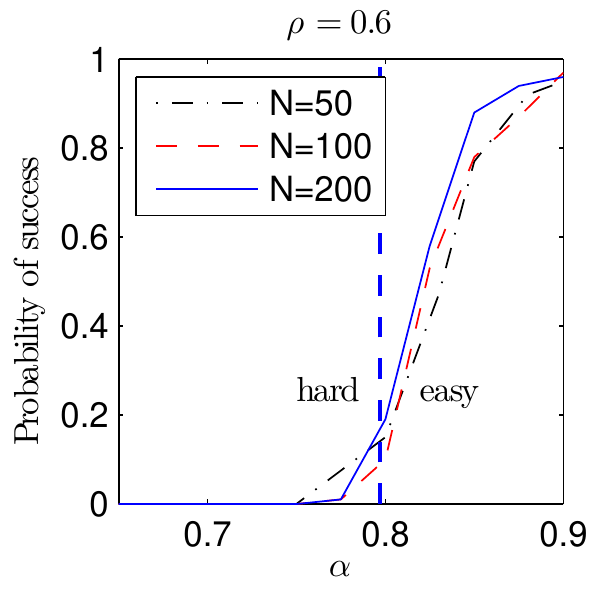}%
}
\caption{Empirical probability of success (defined by ${\rm nMSE}<10^{-6}$), for the experiments presented on figures~\ref{fig:comp01} and~\ref{fig:comp06}.
Due to the finite size, the position of the curves slightly vary for different values of $M$.
Finite size effects allow a fraction of successful instances inside the hard phase for $\rho=0.1$, but much less for $\rho=0.6$.}
\label{fig:PT}
\end{figure}

 Figure~\ref{fig:PT} shows that this effect is quite important for $\rho=0.1$, but nearly inexistent for $\rho=0.6$.
 The reason for this is the much higher nMSE of the unstable fixed point for $\rho=0.1$ than for $\rho=0.6$.

Remember that in \pbig, the initial estimates of $\Uv$ and $\Vv$ are random. 
While in some regions of the phase diagram and with small signal sizes, running the algorithm from several of those random initial estimates might be sufficient, 
in general it would be preferable to have a procedure that systematically produces good initializations.
Previous works stress this fact as well and often rely on an
initialization from spectral methods~\cite{lee2013near,nuclearNormMin,JainLowRank,laffertyLowRank}.
 
Another difference between figures~\ref{fig:PT_01} and~\ref{fig:PT_06} is that in the latter, the algorithm fails for a significant fraction 
of instances inside the ``easy'' phase, which is not the case in the former. 
The fact that the fraction of such failed instances decreases with increasing signal size $\US$ seems to indicate that this is as well a finite size effect.
Unlike the previously examined finite size effect, this one cannot be explained from the state evolution, as it has a unique fixed point in the ``easy'' phase.

\section{Conclusion}

In this paper, we provide an asymptotic analysis of Bayesian low-rank matrix compressed sensing.
We employ the replica method of statistical physics to obtain the so-called state evolution equations,
whose fixed points allow us to determine if inference is easy, hard or
impossible. The state evolution equations describe the behavior of
associated message passing algorithm P-BiG-AMP that was derived and studied
previously in \cite{parker2015parametric}. This work inscribes in a
line of work where approximate message passing was derived and
analyzed on related estimation problems such as compressed sensing \cite{donoho2009message,gamp}, or
matrix factorization \cite{bigamp1,bigamp2,kabashima2014phase}.

An interesting point concerning the state evolution equations is that
they are the same as those for the matrix factorization problem
derived in \cite{kabashima2014phase}. Related observations were made in \cite{Donoho21052013}.

Our analysis, just as the algorithm, is written for a generic
separable prior and output channel.
We analyze in detail the phase diagram for Gaussian noise on the output and Gauss-Bernoulli prior on
both the factors. 
A striking point in the phase diagram is that the $\alpha$
(eq.~(\ref{eq:global_alpha})) needed for the recovery to be tractable does not go
to zero as the factors become very sparse. This is a remarkable
difference between the matrix and the linear compressed sensing.
We show  numerically that there is an excellent
agreement between the theoretical analysis and the performances of the
P-BiG-AMP algorithm.
We observe that for the simulated system sizes, the algorithm performs better
than what could be expected from the asymptotic theoretical analysis. 
However, we explain this as a finite size effect in terms of state evolution fixed points and stress the importance of a good initial estimate 
in order to perform inference outside of the easy phase. Our analysis quantifies
how ``good'' the initialization needs to be for large systems to allow
tractable recovery.

\section*{Acknowledgement}
Philip Schniter's work on this project was supported in part by the National Science Foundation under grant CCF-1527162.
Christophe Schülke's work was supported in part by Universit\'e franco-italienne and in part by the ERC under the
European Union's 7th Framework Programme Grant Agreement 307087-SPARCS.

\appendix
\section{Details for the derivation of the message-passing algorithm}
\label{app:amp}
Here, we complete the derivation of the message-passing algorithm starting with equation~(\ref{eq:messuhNice}):
\begin{align}
 \messuh_{\yi \to \ui \ri}(u_{\ui \ri}) \propto f_0^Y \left( \Zh_{\yi \to \ui \ri} + F_{\yi \ui \ri} u_{\ui \ri} , \Zb_{\yi \to \ui \ri} + H_{\yi \ui \ri} u_{\ui \ri} + G_{\yi \ui \ri} u_{\ui \ri}^2 \right) . 
\end{align}
We first make a Taylor expansion of this message at order $2$ around $u_{\ui \ri}=0$. We drop all indices for this calculation and use simplified notations 
$f = f_0^Y(\Zh, \Zb)$, $\partial_1 = \frac{\partial}{\partial \Zh}$ , $\partial_2 = \frac{\partial}{\partial \Zb}$ :
\begin{align}
 \tilde{m}(u) &\propto f + u \left( F \partial_1 f + H \partial_2 f \right)  \nonumber  \\
				    &+ \frac{1}{2} u^2 \left( F^2 \partial_1^2 f + H^2 \partial_2^2 f + 2 F H \partial_1 \partial_2 f + 2 G \partial_2 f \right) + o(u^2).   \label{eq:messuhExpanded}
\end{align}
We can rewrite $\tilde{m}$ as a Gaussian 
\begin{align}
  \tilde{m}(u) &\propto \NN(u; \hat{p}, \bar{p})  + o(u^2) \label{eq:messuhAsGaussian}
\end{align}
by identifying the coefficients of the Taylor expansion above with the Taylor expansion of a Gaussian
\begin{align}
 \NN(x;\frac{a}{b},-\frac{1}{b}) \propto 1-ax+\frac{b+a^2}{2}x^2 + o(x^2).   \label{eq:gaussianExpanded}
\end{align}
Note that the form~(\ref{eq:messuhAsGaussian}) is only valid around $u=0$: $\tilde{m}$ is not Gaussian. However this form makes calculations easier.
Identification of the coefficients in~(\ref{eq:messuhExpanded}) and~(\ref{eq:gaussianExpanded}) leads to
\begin{align}
  \bar{p} &= -\left[ F^2 \left( \frac{\partial_1^2 f}{f} - \left(\frac{\partial_1 f}{f}\right)^2 \right) + 2 G \frac{\partial_2  f }{f}  \right.  \nonumber   \\
			  &+\left.  H^2 \left( \frac{\partial_2^2 f}{f} - \left( \frac{\partial_2 f}{f} \right)^2 \right) + 2F H \left( \frac{\partial_1 \partial_2 f}{f} -\frac{\partial_1 f}{f} \frac{\partial_2 f}{f} \right) \right]^{-1} \label{eq:pb} \\
  \hat{p} &= - \bar{p} \left( F \frac{\partial_1 f}{f} + H \frac{\partial_2 f}{f} \right)  \label{eq:ph}
\end{align}
We can now treat the $m$-messages from eq.~(\ref{eq:mess_u}). The product is easy to handle as it is a product of Gaussians
\begin{align}
 \prod_{\yibis \neq \yi} \tilde{m}_{\yibis \to \ui \ri} (u_{\ui \ri}) \propto \prod_{\yibis \neq \yi} \NN(u_{\ui \ri}; \hat{p}_{\yibis \to \ui \ri}, \bar{p}_{\yibis \to \ui \ri}) \propto \NN(u_{\ui \ri}; \Uh_{\ui \ri \to \yi}, \Ub_{\ui \ri \to \yi}) , 
\end{align}
which allows us to write
\begin{align}
 \Ub_{\ui \ri \to \yi} &= \left( \sum_{\yibis \neq \yi} \bar{p}_{\yibis \to \ui \ri}^{-1} \right)^{-1} \label{eq:Ub} \\ 
 \Uh_{\ui \ri \to \yi} &= \Ub_{\ui \ri \to \yi} \sum_{\yibis \neq \yi} \left(   \frac{\hat{p}_{\yibis \to \ui \ri}}{\bar{p}_{\yibis \to \ui \ri}}         \right)  \label{eq:Uh}
\end{align}
In the sums above, some of the non-leading order terms stemming from~(\ref{eq:pb},\ref{eq:ph}) have a vanishing contribution in the limit where $(\US,\VS,\YS) \to \infty$ and will therefore be neglected.
The table below analyzes the orders of magnitude and possible signs of all quantities in~(\ref{eq:pb},\ref{eq:ph}). 
In the third and fourth line, we use this to analyze the order of magnitude of a sum of $\YS$ of those terms, as appears in~(\ref{eq:Ub},\ref{eq:Uh}) and what this leads to 
when $\YS \propto \rank \US$, which is the scaling we are interested in.
\begin{center}
\begin{tabular}{|c|c|c|c|c|c|c|}
\hline
  & $F$ & $G$ & $H$ & $F^2$  & $H^2$ & $FH$ \\
  \hline
  scales as: & $\frac{1}{\sqrt{\rank \VS}}$ & $\frac{1}{\rank \VS}$ & $\frac{1}{\rank \sqrt{\US \VS}}$ & $\frac{1}{\rank \VS}$  & $\frac{1}{\rank^2 \US \VS}$ & $\frac{1}{\rank^{3/2} \VS}$  \\
  sign: & $\pm$ & $+$ & $\pm$ & $+$ & $+$ & $\pm$  \\
  sum over $\YS$ & $\frac{\sqrt{\YS}}{\sqrt{\rank \VS}}$ & $\frac{\YS}{\rank \VS}$ & $\frac{\sqrt{\YS}}{\rank \sqrt{\US \VS}}$ & $\frac{\YS}{\rank \VS}$ & $\frac{\YS}{\rank^2 \US \VS}$ & $\frac{\sqrt{\YS}}{\rank^{3/2} \VS}$  \\
  $\YS \propto \rank \VS$ & 1 &1 & $\frac{1}{\sqrt{\rank \US}}$ & 1  & $\frac{1}{\rank \US}$ & $\frac{1}{\rank \sqrt{\VS}}$ \\
  \hline
\end{tabular}
\end{center}
This analysis is based on the fact that:
\begin{itemize}
 \item $\AM$ has random \iid elements of mean 0 and variance $1/(\rank \US \VS)$
 \item $\Uv$, $\Vv$ and $\zv$ have zero-mean elements of order 1, therefore all estimators of type $\uhv,\Uhv$, etc. are of order 1 as well, either positive or negative
 \item variances of type $\ubv, \Ubv$, etc. are positive and of order 1
 \item all quantities of the type $\frac{\partial_i f}{f}$ are of order 1.
\end{itemize}
With the help of the table, we can neglect all terms that have a vanishing contribution.
Furthermore,using the relations~(\ref{eq:fpartial1},\ref{eq:fpartial2}) and the definition of the $g$ functions~(\ref{eq:gh}),
it can be shown that
\begin{align}
\frac{\partial_1 f^Y}{f^Y} &= \gh^Y  , \\
\frac{\partial_1^2 f^Y}{f^Y} - \left( \frac{\partial_1 f^Y}{f^Y} \right)^2 &= \gb^Y , \\
\frac{\partial_2 f^Y}{f^Y} &= \frac{1}{2} \left( \gb^Y - (\gh^Y)^2 \right).
\end{align}
In the end, the resulting expressions for~(\ref{eq:Ub},\ref{eq:Uh}) are given in~(\ref{eq:Ub_final},\ref{eq:Uh_final}).

\section{Details for the replica calculation} 
\label{app:replica}
\paragraph{Covariance matrix of $\mathbf{z}_l$}
We treat $z_{\yi}^a = [\A(\Uv^a (\Vv^a)^{\top})]_{\yi} $ as a random variable of $\A$ and look at the covariance between two of those variables:
\begin{align}
 \langle z_{\yi}^a z_{\yibis}^b \rangle &= \langle \left( \sum_{\ui \vi} \as_{\yi}^{\ui \vi} \sum_{\ri} u_{\ui \ri}^a v_{\vi \ri}^a \right) \left( \sum_{\uibis \vibis} \as_{\yibis}^{\uibis \vibis} \sum_{\ribis} u_{\uibis \ribis}^b v_{\vibis \ribis}^b \right) \rangle  \\
 &= \langle \sum_{\ui \uibis} \sum_{\vi \vibis} \as_{\yi}^{\ui \vi} \as_{\yibis}^{\uibis \vibis} \sum_{\ri \ribis} u_{\ui \ri}^a u_{\uibis \ribis}^b v_{\vi \ri}^a v_{\vibis \ribis}^b \rangle   \\
 &= \sum_{\ui \uibis} \sum_{\vi \vibis} \langle \as_{\yi}^{\ui \vi} \as_{\yibis}^{\uibis \vibis} \rangle  \sum_{\ri \ribis} u_{\ui \ri}^a u_{\uibis \ribis}^b v_{\vi \ri}^a v_{\vibis \ribis}^b
\end{align}
As the elements of $\AM$ are \iid with zero mean and variance $1/(\rank \US \VS)$, 
we have  $\langle \as_{\yi}^{\ui \vi} \as_{\yibis}^{\uibis \vibis} \rangle = \delta_{\yi, \yibis} \delta_{\ui, \uibis} \delta_{\vi, \vibis} \frac{1}{\rank \US \VS}$ 
and thus
\begin{align}
  \langle z_{\yi}^a z_{\yibis}^b \rangle &= \delta_{\yi, \yibis} \frac{1}{\rank \US \VS} \sum_{\ri \ribis} \left( \left( \sum_{\ui} u_{\ui \ri}^a u_{\ui \ribis}^b \right) \left( \sum_{\vi} v_{\vi \ri}^a v_{\vi \ribis}^b \right) \right)  \\
					&= \frac{\delta_{\yi, \yibis}}{\rank} \sum_{\ri \ribis} \left(  \left( \frac{1}{\US} \sum_{\ui} u_{\ui \ri}^a u_{\ui \ribis}^b \right) \left( \frac{1}{\VS} \sum_{\vi} v_{\vi \ri}^a v_{\vi \ribis}^b \right) \right)
\end{align}
We now make the following assumption:
\begin{align}
 \frac{1}{\US} \sum_{\ui} u_{\ui \ri}^a u_{\ui \ribis}^b &= \begin{cases}
                                                Q_u^{ab} = O(1) & {\rm if } \,  \ri=\ribis \\
                                                (Q_u^{ab})_{\ri \ribis} = O(\frac{1}{\sqrt{\US}}) & {\rm if } \, \ri \neq \ribis
                                               \end{cases}		\label{eq:QHyp}
\end{align}
This assumption corresponds to breaking the column-permutation symmetry and more generally the rotational symmetry between
different replicas. We thus assume that the $\ri$-th column of $\Uv^a$ is correlated to the $\ri$-th column of $\Uv^b$ and to none of the others.
We make the same assumption for $\Vv$. Then,
\begin{align}
  \langle z_{\yi}^a z_{\yibis}^b \rangle &= \frac{\delta_{\yi, \yibis}}{\rank} \left(\sum_{\ri} Q_u^{ab} Q_v^{ab} + \sum_{\ri \neq \ribis} (Q_u^{ab})_{\ri \ribis} (Q_v^{ab})_{\ri \ribis} \right).
\end{align}
Due to the hypothesis~(\ref{eq:QHyp}), the second term vanishes, and
\begin{align}
   \langle z_{\yi}^a z_{\yibis}^b \rangle &= \delta_{\ri, \ribis} Q_u^{ab} Q_v^{ab}.
\end{align}
Note that by definition of $Q_u^{ab}$ in~(\ref{eq:QHyp}), $Q_u^{ab}=Q_u^{ba}$. 

\paragraph{Introducing $\Qhu$}
In equation~(\ref{eq:ZwithDeltas}), Dirac $\delta$ functions enforce the relations~(\ref{eq:QHyp}). 
We use the integral representation of these $\delta$ functions to carry on the calculation:
\begin{align}
 \delta\left( \US Q_u^{ab} - \sum_{\ui} u_{\ui \ri}^a u_{\ui \ri}^b \right) &= \frac{1}{2 \pi \imath} \int \dd \tilde{Q}_U^{ab} e^{- \tilde{Q}_U^{ab}\left( \US Q_u^{ab} - \sum_{\ui} u_{\ui \ri}^a u_{\ui \ri}^b \right)}.
\end{align}
The product of all these $\delta$ functions thus gives
\begin{align}
 \prod_{a \leq b} \delta\left( \US Q_u^{ab} - \sum_{\ui} u_{\ui \ri}^a u_{\ui \ri}^b \right) &\propto \int \dd \tilde{\mathbf{Q}}_U e^{- \US \sum_{a \leq b} \tilde{Q}_U^{ab} Q_u^{ab}} e^{ \sum_{\ui} \sum_{a \leq b} \tilde{Q}_U^{ab} u_{\ui \ri}^a u_{\ui \ri}^b}.
\end{align}
Note that because $Q_u^{ab}=Q_u^{ba}$, the replica indices in the sum are $a \leq b$.
Finally, we make a change of variables
\begin{align}
 \forall a, \quad \hat{Q}_U^{aa} &= 2 \tilde{Q}_U^{aa} \\
 \forall (a,b) \text{ with } a\neq b, \quad \hat{Q}_U^{ab} &= 4 \tilde{Q}_U^{ab}
\end{align}
which allows us to obtain the following formulas
\begin{align}
 \sum_{a \leq b} \tilde{Q}_U^{ab} Q_u^{ab} &= \frac{1}{2} \Tr(\Qu \Qhu),    \\
 \sum_{a \leq b} \tilde{Q}_U^{ab} u_{\ui \ri}^a u_{\ui \ri}^b &= \frac{1}{2} \mathbf{u}_{\ui \ri}^{\top} \Qhu \mathbf{u}_{\ui \ri},
\end{align}
where we introduced the vector $\mathbf{u}_{\ui \ri} = (u_{\ui \ri}^0 \dots u_{\ui \ri}^n)^{\top}$. 
We change the integration variable from $\tilde{\Q}_U$ to $\Qhu$, and we obtain the expression~(\ref{eq:ZwithQhats}).

\bibliographystyle{plain}
\bibliography{refs_matrixCS}

\end{document}